\documentclass[preprint]{aastex}
\shorttitle{PDS\,365}
\shortauthors{Drake et al.}

\begin{document}

\title   {RAPIDLY-ROTATING LITHIUM-RICH K GIANTS:
          THE NEW CASE OF THE GIANT PDS\,365}
\author{Natalia A. Drake\altaffilmark {1,2}, Ramiro de la Reza\altaffilmark {1},
        Licio da Silva\altaffilmark {1} \& David L. Lambert\altaffilmark {3}}

\affil{$^{1}$Observat\'orio Nacional - Rio de Janeiro - Brazil}
\affil{$^{2}$Sobolev Astronomical Institute, St. Petersburg State University,
              Russia}
\affil{$^{3}$Department of Astronomy, University of Texas, USA}
\email{drake@on.br}

\begin{abstract}
PDS\,365 is a newly detected rapidly rotating ($v\sin i = 20\,$km\,s$^{-1}$)
single  low mass giant star which   with HD\,233517 and HD\,219025
forms a remarkable ensemble of single K giants with the unique
properties of rapid rotation,  very strong Li lines, an asymmetrical
H${\alpha}$ profile, and a  large far infrared  excess.
Their $v\sin i$ values are between 18 and 23 km\,s$^{-1}$, and  LTE Li
abundances, $\log \varepsilon{\rm (Li)}$, are between 2.9 and 3.9.
Detailed  analysis of PDS\,365 reveals it to be  a  $\sim 1M_\odot$ 
giant with a value of $^{12}$C/$^{13}$C approximately equal to 12.
A clear relation between high rotational velocities
and very high Li abundances for K giant stars is found only when asymmetrical
H${\alpha}$ profiles and large far-infrared excesses are present.
If we consider  single  K giants,
we find that among rapid ($v\sin i \ge 8\, $km s$^{-1}$) rotators,
a very large proportion  ($\sim 50$\%) are  Li-rich giants.
This proportion is in  contrast with a very low   proportion ($\sim 2$\%) of
Li-rich stars among the much more common  slowly rotating K giants.
This striking difference is discussed in terms of proposed mechanisms for
Li enrichment.

\end{abstract}

\keywords{stars: abundances---stars: activity---stars: chemically peculiar---stars:
circumstellar matter---stars: individual (PDS\,365)---stars: mass loss---stars:
rotation}

\section {INTRODUCTION}

It is well known that single K giant stars are very slow rotators having  a mean 
$v\sin i$  value of the order of 2 km\,s$^{-1}$
(De Medeiros, Da Rocha \& Mayor 1996).                         
Rapidly rotating giants are rare; here, we consider a star with
$v\sin i \ge 8$ km\,s$^{-1}$ to be a rapid rotator.
Currently, surveys with higher spectral  resolution than previously used
and new methods for measuring the
$v\sin i$ values are uncovering additional  examples of rapid rotators.
One  case is the giant HD\,37434 with  $v\sin i \simeq$ 65 km\,s$^{-1}$  
(De Medeiros \& Mayor 1999).
Some previous searches of rapidly-rotating giants sought to  test and account for 
a possible relation between high chromospheric activity  and the presence of strong 
Li lines.
Some special surveys were then made in order to try  to find correlations between
rotation, chromospheric activity and Li abundance.
Randich, Gratton, \& Pallavicini (1993) and  Randich, Giampapa, \& Pallavicini (1994)
did not find a correlation between rotation and Li for a group of chromospherically 
active giants.
For a selected group of Li-rich K giants,
De Medeiros, Melo, \& Mayor (1996)                   
found  no correlation between Li and rotation.

De Medeiros et al. (2000) performed  a  statistical analysis  to test the existence 
of a Li -- rotation connection in single giants for stars hotter and cooler than 
G0\,III, the spectral type marking the approximate boundary at which the average 
rotational velocity drops to the low values common among K giants.
A general trend was found in that giants cooler than G0\,III are characterized by 
lower rotation rates and lower Li abundances. 
It is the strong exceptions to this pattern that are the focus of our paper.
For several years, the bright K1\,III star HD\,9746 was the only known
rapidly rotating Li-rich giant ($v\sin i \sim 8\;{\rm km\,s^{-1}}$).
Now approximately twenty rapidly rotating K giants  are known
with varying levels of lithium abundance with several showing remarkable
lithium overabundances relative to abundance expected and observed in
the typical slowly rotating K giant for which the convective
envelope has diluted the lithium abundance by a factor of about 50
relative to its main sequence value.
In Table~1 we  list  all the apparently single K giants stars having 
$v\sin i \ge 8$ km\,s$^{-1}$ together with their spectral types, Li abundances, and
manifestations of activity. Comments on some individual stars are presented
in the Appendix.
These rapid rotators have been selected from the literature and have  spectral types 
later than or equal to G8 (HD\,6665 is an exception, see however the appendix notes
on this star) and luminosity classes III, 
III/II and II. Unfortunately, a Li abundance remains to be obtained for some stars.
The $v\sin i$ values were collected in general from three main sources
(De Medeiros \& Mayor 1999; Fekel \& Balachandran 1993; Fekel 1997) in 
which rotation is measured by  different techniques.
In general, these works give similar values taking into account the
errors involved. The only apparent exception is  the extreme rotator HD\,37434.

In considering  HD\,9746 and other rapidly rotating Li-rich giants, Fekel \& 
Balachandran (1993) suggested a scenario to explain the  relation between Li and 
rotation.
The discovery that the   majority of the Li-strong single giants present large far 
IR excesses (Greg\'orio-Hetem et al. 1992; Greg\'orio-Hetem, Castilho, \& Barbuy 
1993)   
introduced a new parameter -- mass loss -- into the discussion 
(de la Reza, Drake, \& da Silva 1996).        
Following this, two interesting giants were discovered: HD\,233517 
                   (Fekel et al. 1996)         
and HD\,219025
(Fekel \& Watson  1998; Jasniewicz et al. 1999).         
Both are very Li-rich objects but are more rapidly rotating than HD\,9746:
the $v\sin i$ values of HD\,233517 and HD\,219025 are respectively 17.6 
km\,s$^{-1}$
(Balachandran et al. 2000) and 23 km\,s$^{-1}$ (Jasniewicz et al. 1999), 
in contrast to 8.7 km s$^{-1}$ for HD\,9746.

In this paper, we present a new K giant star, PDS\,365, with similar
properties to HD\,233517 and HD\,219025. PDS\,365 was
discovered in the Pico dos Dias Survey (PDS)
(Greg\'orio-Hetem et al. 1992;                          
Torres et al. 1995, Torres et al. 2002; de la Reza et al. 1997).         
In section 2, we present the  main properties of this new object and in
section 3, after inferring new properties of the rapidly rotating 
single K giants, we discuss the problem of Li enrichment.


\section {THE STAR PDS\,365}

\subsection {Observations}


The  star PDS\,365 classified as  K1\,III  (Torres 1998, Torres et al. 2000) is a
faint giant ($V= 13.15\;$mag) and the optical counterpart of  IRAS$\,13313-5838$ 
with RAJ2000 and DECJ2000 equal respectively to ${\rm 13^h 34^m37.6^s}$ and 
$-58^\circ 53' 34''$. It is located near the Galactic plane 
($\ell = 308\fdg 5$, $b = 3\fdg 5$) and, in particular, near a
complex of young stars found by
Efremov \& Sitnik (1988).             
The complex's radial velocity 
relative to the LSR ($-42\;$km\,s$^{-1}$) given by Efremov \& Sitnik, is
almost the same as the star's  velocity.

High dispersion spectra of the star PDS\,365 were obtained in May 1996 with
the 4.0 m telescope at CTIO, Chile,  covering the wavelength interval between
5120 \AA\ and 8370 \AA\  at a spectral resolution of $R=33,500$.
Standard IRAF procedures were used for data reduction.
Very high resolution spectra $(R=181,000)$ of the rapidly rotating 
stars HD\,9746, HD\,233517, and HD\,203251
were obtained with the 2.7 m telescope of the McDonald Observatory.
Nine spectra of HD\,219025 and the spectra of HD\,6665, HD\,129989, HD\,181475,
HD\,284857, $\epsilon$~Vir, and a second spectrum of PDS\,365 were obtained with the
1.52 m telescope at La Silla, Chile,  using the high resolution spectrograph FEROS
under the Observat\'orio Nacional (Brazil)/ESO agreement.
The observation log is given in Table~2.


\subsection {Fundamental  Parameters}

Our analysis of PDS\,365 began by selecting as large a sample as possible of
neutral and ionized iron lines  free of blends.
Line identifications and wavelengths were taken from Moore et al. (1966).
After eliminating lines suspected to be blended, we used  21 Fe\,{\sc i}  and 
3 Fe\,{\sc ii} 
lines for the determination of the fundamental stellar parameters.
The oscillator strengths, $\log gf$, were taken from McWilliam \& Rich (1994)
who considered $gf$-values obtained by different authors and created a
unique system by intercomparison of common  $\log gf$-values.            
Following the usual iterative procedure, we derived the effective  temperature and 
microturbulence by requiring that the iron abundances
be independent of excitation potential and of the equivalent width (EW) (Figure 1).
The surface gravity, $\log g$, was derived from the ionization equilibrium equation 
by finding the value for which the iron abundances from Fe\,{\sc i} and 
Fe\,{\sc ii} coincide.
Table 3 lists the EW measurements, the $\log gf$ values, the low excitation potentials
and final abundances.
We derived the following parameters for PDS\,365: $T_{\rm eff}=4540\,$K, 
$\log g=2.2$, $\xi_{\rm m}= 1.8\,$km\,s$^{-1}$
and ${\rm [Fe/H]}=-0.09$. Line blending due to rapid rotation limits the
number of unblended Fe\,{\sc ii} lines, and, hence, the  accuracy of the $\log g$
determination. 
We estimate $\log g$ is accurate to about 0.3 dex. The synthetic spectrum shown in 
Figure~2  calculated for the spectral region containing Fe\,{\sc ii} line 6084 \AA\ suggests 
that our determination is a good one.
Since the parallax of  PDS\,365 has not been measured, we adopt an absolute
magnitude $M_{V} = 0.6\;$mag, a typical value for K1\,III stars
(Schmidt-Kaler 1982).
With this value, we  estimate an approximate stellar luminosity, mass and radius:
$L=72M_\odot$, $M=1.1M_\odot$ and $R=14R_\odot$ respectively.

To calculate the distance to PDS$\,$365 we used the photometric data obtained during 
the PDS
(Torres et al. 2002):                                      
$V=13.15\;$ and $(B-V)_{\rm obs}=1.50\;$.
The typical value of color index for giants of this temperature is $(B-V)_{0}=1.10\;$
(Schmidt-Kaler 1982),                                           
which results in color excess of $E_{B-V}=0.40\;$.
Thus, the total absorption is $A_V=\Re \cdot E_{B-V}=1.45$,
where $\Re=3.62$ was calculated using the formula from
Schmidt-Kaler (1982).                                               
In this way, we derived a distance of PDS\,365 of $d=1.7\,$kpc.
The total absorption is contributed by the star's circumstellar shell (CS)
and the interstellar medium (IM): $A_V={A_V}^{CS} + {A_V}^{IM}$.
To separate the reddening due to the IM  and CS and estimate an optical depth 
($\tau_V$) of the CS, we analyzed the well defined values of $E_{B-V}$ of a sample of 
open clusters situated close to the line of sight to PDS\,365.
For the IM  reddening, we used a value of $E_{B-V}=0.25$
(Clari\'a 1980)                                                    
determined for the open cluster NGC\,5138 ($\ell=307\fdg 55$, 
$b=+3\fdg 55$) which is situated at the distance of 1.8\,kpc, almost the same as 
PDS\,365.
We thus estimated the absorption in the CS to be ${A_V}^{CS}=$\, 0.55 and
$\tau_V=$0.50.
This inference of circumstellar extinction is qualitatively consistent
with a far IR excess.

The projected rotational velocity ($v\sin i = 20\,$km\,s$^{-1}$) was obtained
by means of a comparison of observed and synthetic spectra.
Synthetic spectra were calculated with different values of
$v\sin i$ with steps of 0.5 km\,s$^{-1}$.
The precision of $v\sin i$ is of the order of 1 km\,s$^{-1}$.
A macroturbulence velocity of 3\,km\,s$^{-1}$ was adopted as a typical value for red
giants (Fekel 1997).                                          

The radial velocity  ($V_{\rm rad} =-38.3\pm 0.5$ km\,s$^{-1}$) was obtained
using telluric H$_2$O lines and O$_2$ lines as the comparison lines.
H$_2$O wavelengths were taken from
Lundstr\"om et al. (1991)                                         
and for O$_2$ from the catalog of
Moore et al. (1966).  
Our recent FEROS spectrum of PDS\,365 provided a new measurement of the radial velocity:
$V_{\rm rad} = -37.4\pm 0.3$ km\,s$^{-1}$ which agrees with the previous result.
An old measurement using however a somewhat lower resolution, gave the value of
$V_{\rm rad} =-39$ km\,s$^{-1}$ (Torres 1998). We  conclude from these
three measurements  that PDS\,365 appears to be a single star.
Determined parameters of PDS\,365 are summarized  in Table~4.


\subsection {Stellar Abundances}     

Using the fundamental parameters determined above, a detailed LTE spectral synthesis 
analysis was done using the current version of the MOOG program
(Sneden 1973) and  Kurucz (1993) atmospheric models                      
to obtain CNO and Li abundances and the $^{12}$C/$^{13}$C ratio.
Due  to the interdependence of C, N, and O abundances caused by the formation of 
molecules, an iterative scheme was implemented to derive the abundances of these 
elements.
The head of the 
C$_2$ 0-1 band of the  Swan 
$d^3\Pi_g - a^3\Pi_u$ system
at 5635 \AA\ was used in the determination of carbon abundance.
The wavelengths of C$_2$ ($^{12}$C$^{12}$C) molecular lines were taken from
Phillips \& Davis (1968). Electronic oscillator strength, $f_{\rm el}=0.33$,
determined in Lambert (1978) was used in our calculations. H\"onl-London factors
were calculated using formulae from Kovacs (1969).
The nitrogen abundance and the carbon isotope ratio $^{12}$C/$^{13}$C were
obtained by comparing the observed and theoretical line profiles for the
$^{12}$CN and $^{13}$CN lines of 2-0 band of the CN red system
$A^2\Pi - X^2\Sigma$ near  8000 \AA. 
The wavelengths for $^{12}$CN lines were taken from the list of
Davis \& Phillips (1963), and those for $^{13}$CN from Wyller (1966).
 Oscillator strength of the band 2-0 of $f_{2-0}=8.4\times10^{-4}$
determined by Sneden \& Lambert  (1982), due to extensive analysis of
molecular lines belonging to different bands of CN red system, were 
used.
The oscillator strengths of $^{13}$CN lines were assumed to be those of the
corresponding lines of $^{12}$CN.
The calculations were performed with the following dissociation potentials:
$D({\rm C_2})=6.15\;$eV, and $D({\rm CN})=7.65\;$eV.
The line lists were tested by fitting  the solar and Arcturus spectra.
The blending of C$_2$ 0-1 lines and the Li\,{\sc i} resonance line with faint CN
lines was taken into account.
Unfortunately, the presence of night sky emission  at [O\,{\sc i}]
6300 \AA\ line prevented our
use of this line to determine the oxygen  abundance.
This abundance was obtained indirectly by adopting the relation [O/H] = 0.5[Fe/H]
(Pagel \& ${\rm Tautvai\check siene}$ 1995) for stars with metallicity 
${\rm [Fe/H] > = -1.0}$.

Results of the abundance analysis of PDS\,365 are shown in Table~4
 (where C$=^{12}$C).
It would be interesting to compare the CNO abundances and $^{12}$C/$^{13}$C ratio for 
PDS\,365 with those corresponding of the other fast rotating Li-rich giants, but 
unfortunately they are generally not known.
 Three exceptions exist: HD\,9746 with $^{12}$C/$^{13}{\rm C}= 28\pm 4$
(Brown et al. 1989),
 and $^{12}$C/$^{13}{\rm C}= 24$ (Berdyugina \& Savanov 1994),  PDS\,100, a high 
rotation Li-rich K giant, with $^{12}$C/$^{13}{\rm C}= 9\pm 1$ (Reddy et al. 2002)
and the Li-poor giant, HD\,112989 with a very low
ratio $^{12}$C/$^{13}{\rm C}= 3.4$ (Tomkin, Luck, \& Lambert 1976).
The low $^{12}$C/$^{13}{\rm C}$ ratio found for PDS\,365  of approximately equal
to 12 (see Fig.~3) constitutes a strong indication that this star is indeed a giant 
star and not a young object.
Other evidence showing that PDS\,365 is not a young object is given by its H$\alpha$ 
absorption line.
Its equivalent width of 1.4\,\AA\, is  far larger than any  EWs of the rare H$\alpha$ 
absorption lines found in pre-main-sequence objects which are mostly found in emission. 
Its H$\alpha$ equivalent width appears to be somewhat larger than the typical value of 
giants corresponding to 1.1 \AA\ (Eaton 1995).


\subsubsection {The CNO abundances}     

  PDS\,365 appears to present anomalous C and N abundances; they differ from those  
expected for a standard giant star that has already passed the first dredge-up 
convective  phase in which $^{12}$C is reduced and $^{14}$N is
increased by a larger factor than those inferred from Table 4.
To confirm this behavior for PDS\,365, we have determined by identical techniques
the CNO abundances for a known standard Li-poor giant ($\epsilon$~Vir). 

We chose $\epsilon$ Vir in part because it was analyzed by
Kj\ae rgaard et al. (1982) and they based the C abundance on the same C$_2$
 5635 \AA\ feature as used in this work for PDS\,365.
Our spectrum of $\epsilon$ Vir was obtained  with FEROS.
We have used similar values of $T_{\rm eff}$, $\log g$, and
microturbulence to those used by Kj\ae rgaard et al. (1982).
The results are presented in Table 5 for which [X/Fe] is the preferred indicator.

Our  C abundance for $\epsilon$~Vir is larger by a factor of 0.10 dex  than 
Kj\ae rgaard et al.'s (1982).
The main reason for this small difference is the treatment of weak blends
on the blue side of the C$_2$  feature at 5635.195 \AA.
Kj\ae rgaard et al. (1982) introduced weak  neutral Fe lines with
excitation potentials around 4.5 eV.
We added   C$_2$ molecular lines with high rotational quantum
numbers from the data bank of Kurucz (1992).

However, the fact that our CNO  abundances for $\epsilon$~Vir appear normal
for a giant star highlights the apparently anomalous C and N abundances of PDS\,365.
In this respect, it is interesting to remark, as  shown in Table~5, that  similar C 
and N anomalies appear also to be present in PDS\,100 and HD\,9746 which are the only 
Li-rich rapidly-rotating giants with CNO values known in the literature.
Curiously, Berdyugina \& Savanov (1994) reported  Li-rich, but slowly-rotating giants 
to have normal C and N abundances. 
Clearly, C, N, and O abundance studies should be extended to larger samples of Li-rich 
giants.


\subsubsection {The Lithium abundances}     

In determining the Li abundance, we consider both the resonance
6708 \AA\ line and the excited line at 6104 \AA. 
 In the spectral synthesis of the
resonance Li line 
we used wavelengths and oscillator strengths from Andersen, Gustafsson \& 
Lambert (1984), and for the secondary line we used the data from Lindg\"ard \&
Nielsen (1977).
For the former line, we
considered the contribution of the isotope $^6$Li in addition to the
usual $^7$Li. 
Three $^6$Li/$^7$Li ratios equal to 0.0, 0.05, and 0.1 were taken into account
and the resulting profiles calculated for the same total Li abundance equal to
$\log\varepsilon{\rm (Li)} =3.3$ are presented in Figure~4. For this Li abundance it 
is clear that the absence of $^6$Li agrees better with observations. 
All attempts to fit the observations with a lesser Li abundance failed. 
In fact, for the case of $^6$Li/$^7$Li$=0.1$ with an abundance of
$\log\varepsilon{\rm (Li)} =2.9$ we can fit the blue
side and part of the center of the resonance line but the red wing will always be much
too wide to fit this part of the line.
Because the weak excited 6104 \AA\ line is
insensitive to the isotopic mix, we can conclude from our observations of the
resonance line that $^6$Li is absent. Figure~5 shows the synthetic (pure $^7$Li) 
and observed spectra for the two lines. The same Li abundance, 
$\log\varepsilon{\rm (Li)} =3.3$ fits both observed profiles.

Due to the star's large rotational velocity, the blend at the blue side 
of the Li\,{\sc i} line at 6104 \AA\  is complex.
We have been unable to match well  the observed spectrum
even using the most recent atomic data (VALD) on the blending  Fe\,{\sc i} and 
Ca\,{\sc i} lines.
In any case, the Li abundance obtained from this line, even with a small error due to 
this complex blend, is compatible with the Li abundance obtained from the resonance 
Li\,{\sc i} at 6708 \AA.

Non-LTE effects may affect the two Li lines differently.
A first approximation to  these effects can be estimated using calculations
by Carlsson et al. (1994).
For the effective temperature and gravity of PDS\,365,  Carlsson et al.'s
corrections are  $-0.17$ for the resonance line and +0.21 for the secondary line. 
Nominally, this implies slightly different abundances from the two lines.
Other non-LTE calculations include the effect of an overlying chromosphere 
(de la Reza \& da Silva 1995), which may affect the sign and
magnitude of the non-LTE calculations. It is clear, however, that
the lithium enhancement of PDS\,365 is only marginally affected by
non-LTE effects.


\subsection {Evidences of Chromospheric Activity and Mass Loss}   

PDS\,365 has a large far-IR excess and a very asymmetrical H$\alpha$ profile.
These properties are similar to those found in the three other  rapidly
rotating Li-rich giants discussed in this work.
The H$\alpha$ line of PDS\,365 and other rapidly rotating giants is shown in Fig.~6.
The core of the line is blue-shifted relative to the rest frame of the star, and the 
blue wing is weakly in emission.
One of the H$\alpha$ profiles of PDS\,365 is very similar to that of HDE\,233517 
(Fig.~6).
The core and wings of the H$\alpha$ profiles of these two stars resemble those
of late-type supergiants (Mallik 1993; Eaton 1995) with chromospheric winds.
Other phenomena indicative of even  more energetic activity include the detection of 
X-rays and optical flares (Konstantinova-Antova \& Antov 2000) from some of 
these stars (see Table 1).
In this study, we  consider a variable H$\alpha$ line as a  manifestation of activity 
in these rapidly-rotating giants (see Fig.~6). The most striking case is that of 
HD\,219025 with a variable central  emission feature, and   blue-shifted  absorption
in the core which varies even on a day-to-day basis (de la Reza et al. in preparation).
These H$\alpha$ profiles may be produced by localized active regions on the stellar 
surface and modulated by rotation.
The blue asymmetry is likely a signature of mass loss.

The signature of  mass loss can be found in other spectral features (Na D lines) and 
in the far-IR excesses. Indeed, the Li-strong giants such as HD\,233517 and HD\,219025 
present satellite absorption features in the blue wings of the resonance Na D lines.
In a  very high resolution spectrogram of HD\,233517 obtained with the 2.7 m McDonald 
telescope, we detected several Na satellite lines; Balachandran et al. (2000)
reported variations between the stellar and satellite lines.
 In the case of HD\,219025, only one strong blue-shifted absorption was detected, but 
the variability of this absorption revealed by several spectra shows clearly that this
line is not of interstellar origin.

Another way to investigate  past mass loss episodes involves  far-IR excesses
measured by IRAS.
Two color-color diagrams of IRAS fluxes are presented in Figures 7a and 7b for K
giants in the IRAS catalog including the faint source extension.
In Fig.~7a are plotted the rapid rotating K giants of Table 1. Fig.~7b which is 
similar to Fig. 1 of de la Reza et al. (1997) contains slowly rotating K giants. 
We must note, however, that for some stars rotational velocities remain to be measured.
In both diagrams, three different regions labeled as I, II, and III are marked.
Region I is defined by the photospheric colors, i.e., no far-IR excesses are present.
The very large majority of Li-poor K giants belong to this region.
Region II is the region where far IR excesses at 25 and 60 microns are pronounced.
 Region III is characterized by an excess at 60 microns.
A scenario connecting a sudden $^7$Li enrichment followed by an ejection
of a detached CS and a subsequent $^7$Li depletion has been proposed by 
de la Reza, Drake \& da Silva (1996) and 
de la Reza et al. (1997) to explain the presence of K giants (rich or poor in Li) in 
these three regions.
In this scenario, a complete CS ejection is represented by a loop beginning in region I,
crossing regions II and III, and finishing again in region I.

Of the Li-rich ($\log \varepsilon$(Li) $\geq 1.5$)  rapidly rotating stars,
none falls in region I, the location of the vast majority of the slowly
rotating and Li-poor giants. Considering the total of  18 rapidly
rotating giants examined for lithium  and  plotted in Figure~7a,
4 of 18 are in region I but  none are Li-rich.
Three more  of the 18 stars are in region II and  each is
extremely Li-rich. Region III hosts 11 stars with known Li abundances 
of which 6 are Li-rich.
The Li-rich and very active single giant BD+70\,959 (ET Dra) (Ambruster 
et al. 1997)  has not
been included in Fig.~7a because the association between the star and an
IRAS source is uncertain. Stars HD\,232862 and HD\,185958 are also not 
represented in Fig.~7a. 
The first one, appears not to be an IRAS source and the second one has 
only upper limits of its fluxes at 25 and 60 microns and 
its corresponding point is out of the figure.
For slowly rotating giants (Figure 7b), the vast majority are 
in region I with a  very low percentage of Li-rich stars.
For region II, 16 of 34 stars are Li-rich. 
There are 68 stars in region III of which 11  are Li-rich.
Several stars lack a Li abundance determination. If these are excluded,
the percentage of Li-rich stars increases slightly.
Perhaps, the most noticeable statistics are the relative lack of rapidly rotating 
giants in region I, and the increased frequency of Li-rich rapidly rotating
giants in regions II and III.
 Considering the 20 rapidly-rotating giants examined for Li in Table 1 
(excluding BD+25\,4819, possibly not a K giant, and HD\,284857, not a rapid rotator),
there are 10 Li-rich giants.
This represents a very large ($\sim 50$\%) fraction of Li-rich giants if compared to the
typical low proportion ($\sim 2$\%) (Brown et al. 1989) 
of Li-rich giants among the low rotators.
These statistics encourage the speculation that the lithium enrichment is common 
among rapidly-rotating K giants.

In summary, we suggest four important (tentative) conclusions:
 1) There is not a   one to one relation between rotation and Li abundances.
 2) Rapid rotation appears to be confined  to spectral types between G8
and K2. This is probably a selection effect arising from the very large
number of early type K giants in common catalogs.
 3) An association of rapid rotation with a lithium enrichment appears
when a high  far IR excess is present.  HD\,9746 is an exception.
 4) As noted above, Li enhancement is particularly common among
rapidly rotating K giants but rare among slowly rotating giants.
The first two conclusions have been found by other authors, but the last two 
are a new result of this work.


\section {Lithium Enrichment of Giants}   

Single K giants are predominantly Li-poor, slowly rotating, and lacking
a far IR excess.
The baseline for lithium abundance in a K giant is set by the
Li abundance that results from dilution arising from the
giant's convective envelope. This may be put at about 
$\log \varepsilon$(Li)
$\leq$ 1.5, a limit that allows for depletion of lithium by the main
sequence progenitor as well as a deeper than predicted convective
envelope. Stars with a Li abundance in excess of this
limit are declared to be Li-rich (see also Charbonnel \& Balachandran 
2000).
Outstanding cases like PDS\,365 are
remarkably Li-rich with a Li abundance approaching or exceeding that
expected of the star-forming cloud.

K giants that are rapidly rotating or Li-rich or having a far IR 
excesses are few in number.
Often, the defining characteristics of these exceptional stars are
correlated. When correlated, the implication must be that either
a single process or related string of processes led to the
anomalies.
An imperfect correlation between excess rotation, lithium, and 
circumstellar
material  does not  necessarily imply unrelated processes but
rather may indicate different times and
timescales for either the appearance or disappearance
of the three  exceptional observational characteristics.

These characteristics promote the following questions:
\begin{itemize}
\item
Rapid rotation: Was this the result of the inhibition of the mechanism
that reduces the surface angular rotation rate of most K giants? Or was
angular momentum added to the envelope from the interior (e.g., a rapidly
rotating core) or the exterior (e.g., a planet was swallowed)?
\item
Lithium: Was the lithium dilution experienced by normal K giants inhibited? 
Or was fresh lithium added to the envelope from the interior (e.g., $^3$He was 
converted to $^7$Li by the Cameron-Fowler [1971] mechanism), the surface 
(e.g., lithium production by stellar flares),   the exterior (e.g.,  planets or brown 
dwarves were swallowed), or through a combination of interior, surface, and 
exterior processes?
\item
Far-infrared excess: Is the circumstellar material a residue of
a protostellar disk/nebula or a dissolving planetary system? Or was the
material ejected by the red giant or its immediate progenitor?
\end{itemize}

Presently, one scenario  appears to account quite well for the existence in low numbers
of rapidly rotating K giants, some of whom may be Li-rich and/or show a far IR excess. 
This is the idea that K giants may swallow a giant planet or brown dwarf. 
An idea, which in connection with lithium enrichment was first mooted by
Alexander (1967), and was recently developed in some quantitative detail by
Siess \& Livio (1999a,b). Accretion and ingestion of giant planets or
a brown dwarf by a red giant is shown by Siess \& Livio (1999b) to account
for a high Li abundance, a far IR excess, a rapid rotation, and X-ray emission in
a few per cent of K giants, where the latter frequency is based in part on the
observed frequency of low mass main sequence stars with planets.

 In this scenario, the stellar lithium abundance saturates when the
mass of accreted material approaches the mass of the convective
envelope. Since the maximum abundance inferred for interstellar
material and young main sequence stars is $\log\varepsilon$(Li) $\simeq
3.0 - 3.3$, one expects this to be the limit for stars that accrete
interstellar-like material with   an undepleted lithium abundance.
This limit will not be reached, however, by accreting one or two
giant planets; the Jovian mass is but one-thousandth of a solar mass.
Accretion of 10 Jovian masses might raise the abundance to
$\log\varepsilon$(Li) $\sim$ 1.5 (Siess \& Livio 1999b). Accretion of a
 brown dwarf with $M \sim 0.1M_\odot$ would put the abundance close to
the undepleted value, a value observed for PDS\,365. Lithium abundances
exceed the undepleted value in several cases; for example,
Balachandran et al. (2000) obtain $\log\varepsilon$(Li) = 4.2 for HDE\,233517. 
Accretion of terrestrial planets can, in principle, increase the
lithium abundance above the initial/interstellar value but only if
the mass of material from which those planets formed greatly exceeds
the mass of the giant's convective envelope. Alternatively,
we note that Denissenkov \& Weiss (2000, also Siess \& Livio 1999b)
propose that the accretion process triggers production of $^7$Li from $^3$He
(see below) to provide additional lithium enrichment at the surface.

Charbonnel \& Balachandran (2000) used Hipparcos parallaxes to place
Li-rich giants on a Hertzsprung-Russell diagram, and to identify
two favored luminosities for Li-rich stars. One  sample including
HD\,9746, HDE\,233517, and HD\,219025 are at the `RGB bump' where
models show that the H-burning shell of low mass red giants burns through
the molecular weight discontinuity left from the initial growth of the
convective envelope that diluted the surface lithium and the 
$^{12}$C/$^{13}$C
ratio. (Intermediate mass Li-rich giants  were found at a higher preferred
luminosity associated with a molecular weight discontinuity predicted
for the interior of an early-AGB star.) A trigger for Li-production
was not suggested but the source of $^7$Li was supposed to be $^3$He, a
residue of the primordial $^3$He and a product from
partial operation of the $pp$-chain in the main sequence
star.

Palacios, Charbonnel, \& Forestini (2001) propose a trigger mechanism
for the RGB-bump stars.
Low mass stars at the luminosity of  RGB bump burn $^3$He to $^7$Be
(and then to $^4$He)
just exterior to the H-burning shell
but, as this is  below the base of the convective envelope, the
surface is not enriched in $^7$Li, the product of electron capture on
$^7$Be. Palacios et al. invoke  efficient diffusion of $^7$Be into the
region between the top of the H-burning shell and the base of the
convective envelope. A consequence is a significant release of
energy in this region primarily from $^7$Li$(p,\alpha)\alpha$ with
the $^7$Li formed from the $^7$Be. This Li-flash turns the region
convective and it merges with the convective envelope enabling
the surface $^7$Li abundance to rise. Subsequent to the Li-flash,
the surface Li is reduced because $^7$Li is destroyed at the base
of the convective envelope.
Palacios et al. complete the  connection between their
theoretical account of Li-enrichment
of red giants and the observations by two assumptions:
(i) the required efficient diffusion is assumed
related to rapid rotation of the core, and (ii) the luminosity increase
from the Li-flash is assumed to result in mass-loss and
formation of a dust shell.

Other proposals linking $^7$Li to the reaction chain
$^3$He$(\alpha,\gamma)^7$Be$(e^-,\nu)^7$Li have been advanced but
physical triggers for initiating the chain have not been
identified (Fekel 1988; Fekel \& Balachandran 1993; de la Reza, Drake,
\& da Silva 1996; de la Reza et al. 1997; Sackmann \& Boothroyd 1999,
2000). In addition, the observational requirement for a coupling of
lithium production to rotation and mass loss has been either overlooked
or subject to nothing more than speculation.

Observations will likely decide whether the lithium of a Li-rich giant
was synthesized internally or gathered by accretion.
Differences
in composition may be used to distinguish between the two
hypotheses. Accretion or ingestion increases not only the lithium
abundance but also the Be and B abundances, two light elements whose
abundances were also reduced by the giant's convective envelope.
Castilho et al. (1999) found Be underabundant in two Li-rich giants
(HD\,787 and HD\,146850)
and concluded that lithium had been produced internally. 
 Neither giant is a rapid rotator.
Comparisons of the C, N, and O isotopic and elemental abundances for 
Li-rich giants with those of
normal giants are likely to prove instructive.

An obvious additional test involves the lithium isotopic ratio. Internal
production implies the surface lithium should be pure $^7$Li. Ingestion
of a giant planet adds $^7$Li and $^6$Li with an abundance ratio presumably
equal to that of the star's natal  interstellar cloud. One supposes that
this ratio is similar to the meteoritic ratio ($^7$Li/$^6$Li = 12) and 
the local interstellar ratio, which is generally similar to the
meteoritic ratio but one gross exception is known (Knauth et al. 2000).
 Our analysis of the 6707 \AA\ feature in PDS\,365 showed $^6$Li to be 
absent. While positive
detection of $^6$Li would certainly have supported the hypothesis of
ingestion, the failure to detect $^6$Li does not necessarily require
rejection of the hypothesis because even mild internal destruction of
lithium following ingestion will lead to very severe losses of $^6$Li
because its destruction by protons occurs 70 times faster than the
destruction of $^7$Li. Additionally, if ingestion serves as a trigger
for $^7$Li production from $^3$He, the $^6$Li abundance may remain
low and even undetectable.

Three other Li-rich giants fail to show $^6$Li. Reddy et al.
(2002) from 6707 \AA\  line profile analysis for a giant (PDS\,100) with
$\log\varepsilon$(Li) = 2.5 found no $^6$Li. Balachandran et al. (2000)
excluded the presence of $^6$Li at (or above) the meteoritic
abundance in the Li-rich giants HD\,9746 and HDE\,233517 not from
line profile analysis but on the grounds that the  
abundance derived from the very strong 6707 \AA\ resonance doublet and the
weaker 6103 \AA\ secondary line were greatly different when $^6$Li was included
in the analysis. With an isotopic shift of 0.15 \AA,
the presence of $^6$Li reduces the abundance needed to fit the equivalent
width of the saturated 6707 \AA\ line.  Applied in this way, the test for
$^6$Li is not definitive because there are other ways to change the
equivalent width of strong lines: e.g., modification of the 
microturbulent
velocity field assumed to be isotropic and depth independent,
changes to the outer layers of the model atmosphere, alternative model
atmospheres,
non-LTE effects. A new attempt to measure the $^6$Li concentration
should be made, with a much higher resolution, as used in this work,
using the profile of the Li\,{\sc i} 6707 \AA\ line along with
a parallel analysis of other strong lines of atom of low
ionization potential, e.g., the Na\,D lines, the K\,{\sc i} 7699 \AA\ line 
and possibly the Ca\,{\sc i} lines. Although the initial searches
for $^6$Li have been unsuccessful, a single detection would be an
enormous stimulant to the search for the mechanism responsible for
lithium enrichment of these red giants.
  

\section {Conclusions}     

In this paper, we report on the discovery of a new rapidly rotating Li-rich
K giant PDS\,365, and discuss some general properties of an ensemble of high 
rotating K giants with $v \sin i \ge 8\;$km\,s$^{-1}$. 
There is a correlation between rapid rotation, mass loss as
measured by the far IR excesses  and an asymmetric H$\alpha$ line, and a 
high Li abundance.  
At present, one theoretical explanation explains in broad terms
these connections: ingestion of material by the red giant provides
lithium enrichment, increases the angular momentum of the giant's
envelope, and stimulates mass loss. Additional lithium enrichment may
occur because the accreted material in  penetrating deep into the convective
envelope to just above the hydrogen burning shell may trigger conversion 
of $^3$He to $^7$Li. 

Red giants harbor an adequate supply of $^3$He from which to synthesize
$^7$Li by the Cameron-Fowler mechanism. Palacios et al. (2001) have
suggested that the Li-flash may operate in giants at the RGB bump, and
speculated on a connection between the efficacy of the flash and
rotation.
This scenario not only agrees with the sudden $^7$Li-enrichment -- mass 
loss connection proposed by de la Reza
et al. (1996, 1997) but, also, supports and enhances the necessity of
considering stellar rotation in understanding
the Li phenomena and this is the main subject of this paper.
Observational tests involving the determination of the $^6$Li, Be, and B
abundances should serve to distinguish stars that gained lithium
by ingestion from those that synthesized it internally.

Nature is so rich in phenomena that one should not insist that  all Li-rich
giants be traceable to a single origin. 
Three origins presently compete for observational approval: ingestion
of material (brown dwarf, giant planets, and terrestrial planets),
ingestion followed by triggered production of $^7$Li from the stellar
reservoir of $^3$He, and internal activation of the $^3$He reservoir.
PDS\,365 would appear to have a lithium abundance in excess of that
directly obtainable by ingestion.

\acknowledgments
N.A.D. thanks FAPERJ for the financial support under the grants 
E-26/151.172/98 and  E-26/171.647/99 and L.\,da\,S. thanks the 
CNPq for grant 200580/97-0. 
DLL acknowledges the support of the Robert A. Welch Foundation of
Houston, Texas.
We thanks the referee for important suggestions that improved the 
presentation of this paper.
SIMBAD and VALD services have been used in this work.
\clearpage

\appendix
\section {Comments on some individual stars.}

{\bf BD+25\,4819}. We have recently observed this star spectroscopically.
Its spectrum appears to correspond to a star hotter than K0 II, the spectral 
type given in the literature.
Its H$\alpha$ EW, equal to 3.2 \AA\ is much larger than  a typical EW of a 
G5 - M5 giant of 1.1 \AA\ (Eaton 1995).
Li lines are undetectable.

{\bf BD+31\,2471}. 
No studies of this star in the Li spectral region are known in the literature.

{\bf HD\,6665}. This Li-rich giant was found  by Strassmeier et al. (2000) 
who give a projected rotational velocity $v\sin i =10.0$ km\,s$^{-1}$
and 7.9 km\,s$^{-1}$. 
We found  a lower value of $v\sin i =6.0$ km\,s$^{-1}$ which includes the  
macroturbulent velocity.
Using the stellar temperature $T_{\rm eff}=4500$\,K
 given  by Strassmeier et al. (2000) and adopting  $\log g=2.00$ and 
[Fe/H]=0.0 we obtain a LTE
Li abundance of $\log \varepsilon{\rm (Li)}=2.7$.
A detailed analysis of this star will be presented later.
Strassmeier et al. (2000) considered this star as G5\,III nevertheless,
a later type would be more consistent with its observed $B - V$ 
value of 1.19. 
However, the $B - V$ value of this star, used for the $T_{\rm eff}$ determination,
may be affected by the absorption of a circumstellar shell. The use of a
somewhat higher temperature will enhance the Li abundance.

{\bf HD\,9746\,=\,OP And} is a well-known Li-rich giant discovered by 
Brown et al. (1989) that has been studied by several authors.
This star has rather high C/N ratio equal to 3.89 (Berdyugina \& Savanov 1994) and
$^{12}$C/$^{13}{\rm C}=28\pm 4$ (Brown et al. 1989).
For stellar parameters proposed by Brown et al. (1989) 
$(T_{\rm eff}$, $\log g$, [Fe/H]) equal to (4420, 2.3, $-0.13$) which are similar
to those used  by Pavlenko et al. (1999) and Balachandran et al. (2000),
we derived an LTE Li abundance value $\log \varepsilon ({\rm Li}) =3.5$
using both resonance and subordinate Li lines, which is in good agreement
with the value  $\log \varepsilon ({\rm Li}) =3.6$ obtained by Pavlenko et al. (1999)
with  allowance for non-LTE effects.
 A photometric period of 76 days has been proposed by Strassmeier \& Hall (1988).
We note the high level of activity as indicated by very strong single-peaked
Ca\,{\sc ii} H and K emission lines (Smith \& Shetrone 2000) and X-ray
emission. This star was found to flare by Konstantinova-Antova \& Antov (2000).


{\bf HD\,31993}.
Three spectra of this star obtained on successive nights show
variations of the  H$\alpha$ line's EW
(1.56 \AA, 1.24 \AA, and 1.18 \AA). 
The photometric period is 28 days (Hooten \& Hall 1990,
Strassmeier et al. 1997).

{\bf HD\,33798}.
This low mass star ($M=1.8M_\odot$, Gondoin 1999) was studied by
Fekel \& Marshall (1991). They found no periodic
radial velocity variation and suggested that the star is single.
Magnitude variations with the period of $P=9.825$ days were found 
(Hooten \& Hall 1990),
which with $v\sin i =29$ km\,s$^{-1}$ measured by Fekel \& Marshall 
(1991)
results in $R\sin i =(5.6\pm 0.4)R_\odot$.
The Hipparcos parallax
of $\pi=8.94\pm 1.35$ mas  results in $M_V=1.68$, $R=6.7R_\odot$,
$i=57^\circ$, and an equatorial rotation velocity $v_{\rm eq} =35\;{\rm km\,s^{-1}}$.
Gondoin (1999), due to its high X-ray luminosity
($L_X=1074.7 \cdot 10^{27}\, {\rm erg\,s^{-1}}$),
considers this star to be a FK Comae candidate. 
Hipparcos observations revealed a faint companion star with
angular separation of $0.358 \pm 0.011$ arcsec, and $\Delta V = 1.85\pm 0.08$
mag.
Flare activity on this star have been detected by Konstantinova-Antova, Antov, \&
Bachev (2000).



{\bf HD\,39152}.
No information on the Li spectral region has been found in the literature.

{\bf HD\,68776}. Small-amplitude variable star ($\Delta V=0.145$ mag,
Percy et al. 1994).
Adopting the stellar parameters $(T_{\rm eff}$, $\log g$, [Fe/H]) equal to 
(4750, 2.5, 0.00) we derived an LTE Li abundance value of
$\log \varepsilon ({\rm Li}) =1.1$. This star has a symmetrical H$\alpha$
profile with an EW equal to 1.26 \AA. Ca\,{\sc ii} emission is absent.
 
{\bf HD\,112989 = 37 Com}.
This intermediate mass star ($M=5.0M_\odot$, De Medeiros et al. 1999) has
the very low ratio $^{12}$C/$^{13}$C=3.4,  equal to the CN-cycle's
equilibrium value (Tomkin et al. 1976).
The low carbon/nitrogen  ratio C/N=0.8
(Lambert \& Ries 1981) also points to a strong mixing undergone by this star.
De Medeiros et al. (1999)
found  significant Ca\,{\sc ii} H and K emission  variability.
Henry et al. (2000) found a periodicity of $P=70$ days, and 
$i=17^\circ$
and noted that the star may undergo radial pulsations.

{\bf HD\,129989}.
Moderate X-ray luminosity $L_X \le 5.1 \cdot 10^{28}\;{\rm erg\,s^{-1}}$,
intermediate mass, $M=4.6M_\odot$, star (Gondoin 1999).
 A recent high resolution spectrum of this star
obtained in this work shows the Li line to be absent.

{\bf HD\,145001}.
A high X-ray luminosity ($L_X = 3.318 \cdot 10^{30}\;{\rm erg\,s^{-1}}$)
intermediate mass star ($M=3.4M_\odot$, Gondoin 1999)


{\bf HD\,181475}. Henry, Fekel, \& Hall (1995) found a variability of
$\Delta V = 0.06$ mag with a periodicity  of $P= 31\pm 1$ days
and presumed that this  is due to pulsations.
Hipparcos data confirmed this variability giving an
intrinsic variability amplitude of $0.055 \pm 0.013$ mag.
A recent spectrum of this star obtained in this work presents no Li lines.

{\bf HD\,185958}.
CN-strong star  (C/O=0.50, Mishenina \& Tsymbal 1997).

{\bf HD\,199098}. No studies are known of this star in the Li 
spectral region.

{\bf HD\,203251}.  For this rapid rotator we found LTE Li abundance equal to 
$\log \varepsilon ({\rm Li}) = 1.5$ similar to that presented by Fekel \& Balachandran 
(1993)
(1.4) using the following stellar parameters $(T_{\rm eff}$, $\log g$, [Fe/H]) 
equal to (4500, 3.00, $-0.3$).

{\bf HD\,219025}. Even if Houk \& Cowley (1975) classified this star as  K2\,IIIp, other
authors, as Whitelock et al. (1995) and Randich et al. (1993) suggested that this
star could be a pre-main sequence star. Also a RS CVn-type class was proposed
for this star by  Cutispoto (1995). Fekel et al. (1996) gave convincing arguments
indicating that this star is in reality a giant star with a radius of $18.2R_\odot$.
Our large number of spectra obtained for this star (see Table 2) do not show the
presence of duplicity. A rotational velocity of $23 \pm 3$ km\,$s^{-1}$ was 
measured by Jasniewicz et al. (1999).  
With  stellar parameters, similar to those proposed by Jasniewicz et al.  
$(T_{\rm eff}$, $\log g$, [Fe/H]) equal to (4500, 2.0, $-0.1$) we deduce a Li abundance 
of $\log \varepsilon({\rm Li}) = 2.9$.
Fekel \& Watson (1998) obtained a larger LTE Li abundance of 3.3 using a hotter
atmosphere ($T_{\rm eff} = 4640\,$K) and a larger Li EW. As mentioned in section 2.4 due
to its interesting properties of variability this star will be the subject of a
subsequent separate work (de la Reza et al. in preparation).

{\bf HD\,232862}.
No information on the Li spectral region has been found in the literature.

{\bf HD\,233517}. 
 This star has been considered in the recent past a main sequence Vega-type star
(Skinner et al. 1995). However, Fekel et al. (1996) showed that this star is not 
only a single giant but also a very Li-rich star. 
Drake (1998) found a LTE Li abundance of $\log \varepsilon({\rm Li}) = 3.9$  
using the following stellar parameters 
$(T_{\rm eff}$, $\log g$, [Fe/H]) equal to (4470, 1.42, $-0.45$), similar,
except the gravity, to those proposed by
Balachandran et al. (2000).

{\bf HD\,284857}.  This is a relatively unknown star in the literature.
We calculated a Li abundance of $\log \varepsilon{\rm (Li)}$ $\simeq 0.3$
adopting the following stellar parameters:
$(T_{\rm eff}$, $\log g$, [Fe/H]) equal to (4250, 1.5, 0.00)
In any case, this giant cannot be considered a rapidly-rotating star. 
Our measured
 $v\sin i$ value is only 4.5 ${\rm km\,s^{-1}}$.
 The equivalent width of the symmetrical H$\alpha$ line is 1.4 \AA.

\clearpage

\figcaption[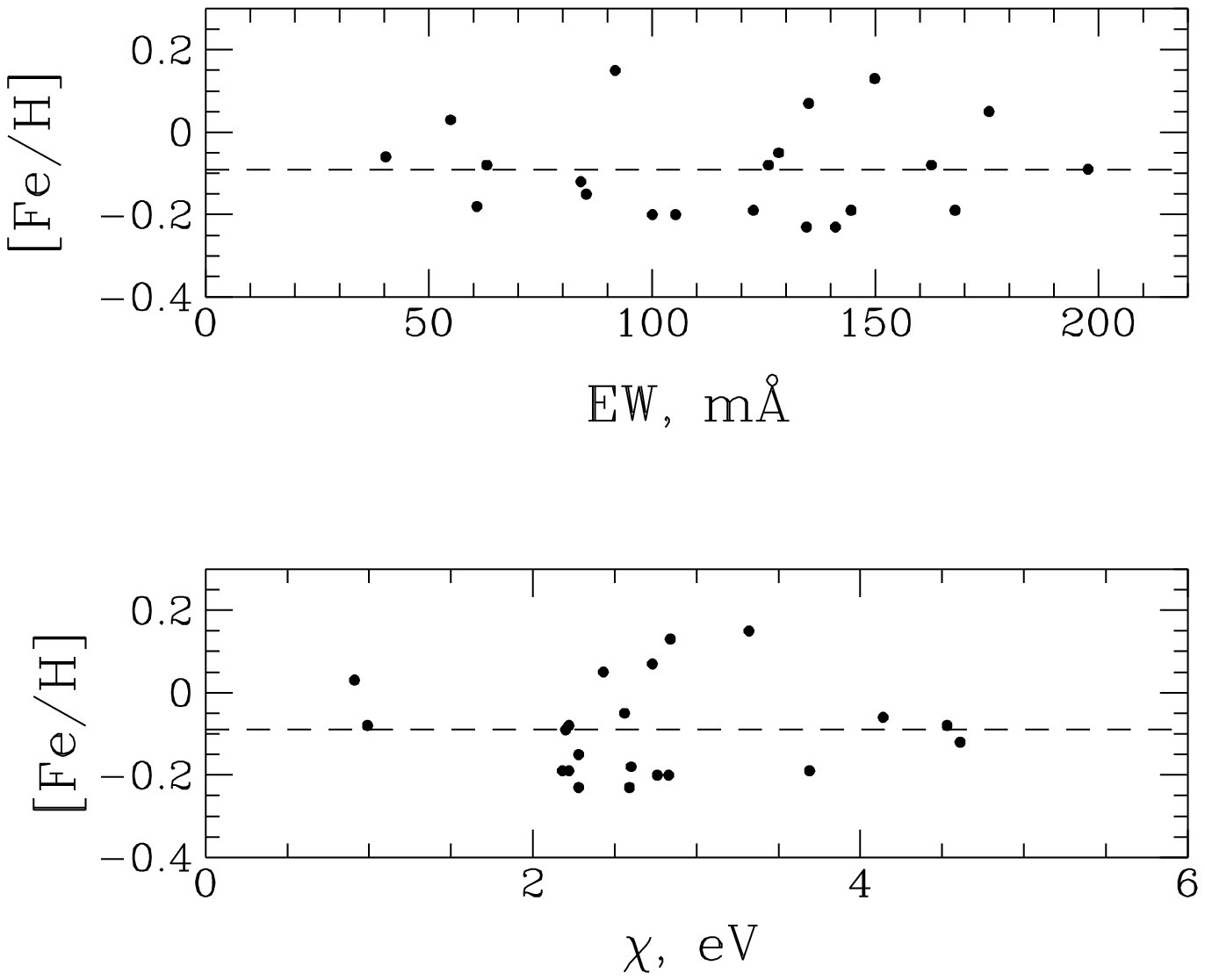]
{Star PDS\,365. Top half: Iron abundances from Fe\,{\sc i} lines {\it vs} equivalent width.
            Bottom half: Iron abundances from Fe\,{\sc i} lines {\it vs} 
excitation potential.}

\figcaption[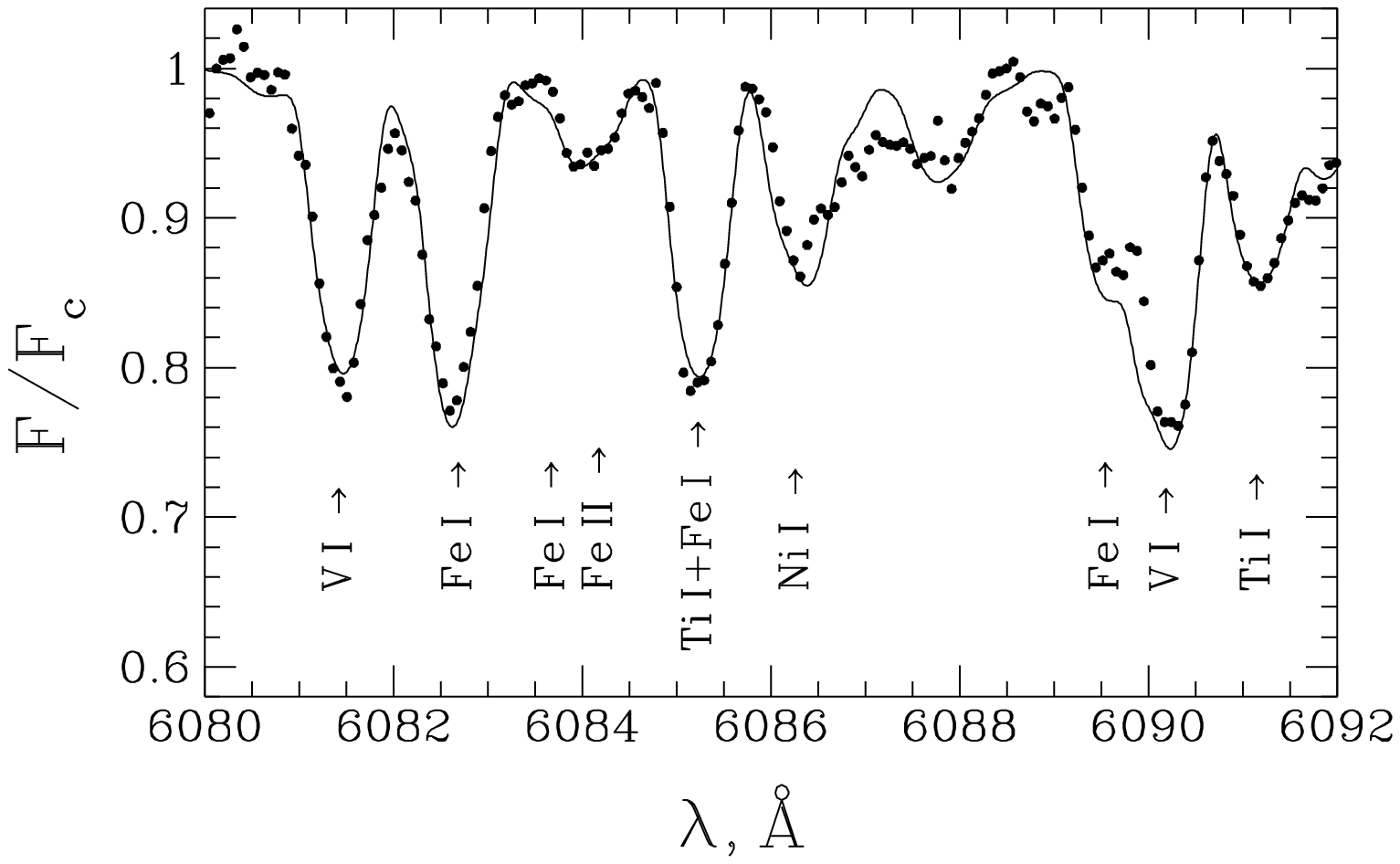]
{Star PDS\,365. The spectral region containing the Fe\,{\sc ii} line at 6084 \AA. 
Observations are represented by separated points. The spectrum synthesis is
made for the stellar parameters presented in Table~4.}

\figcaption[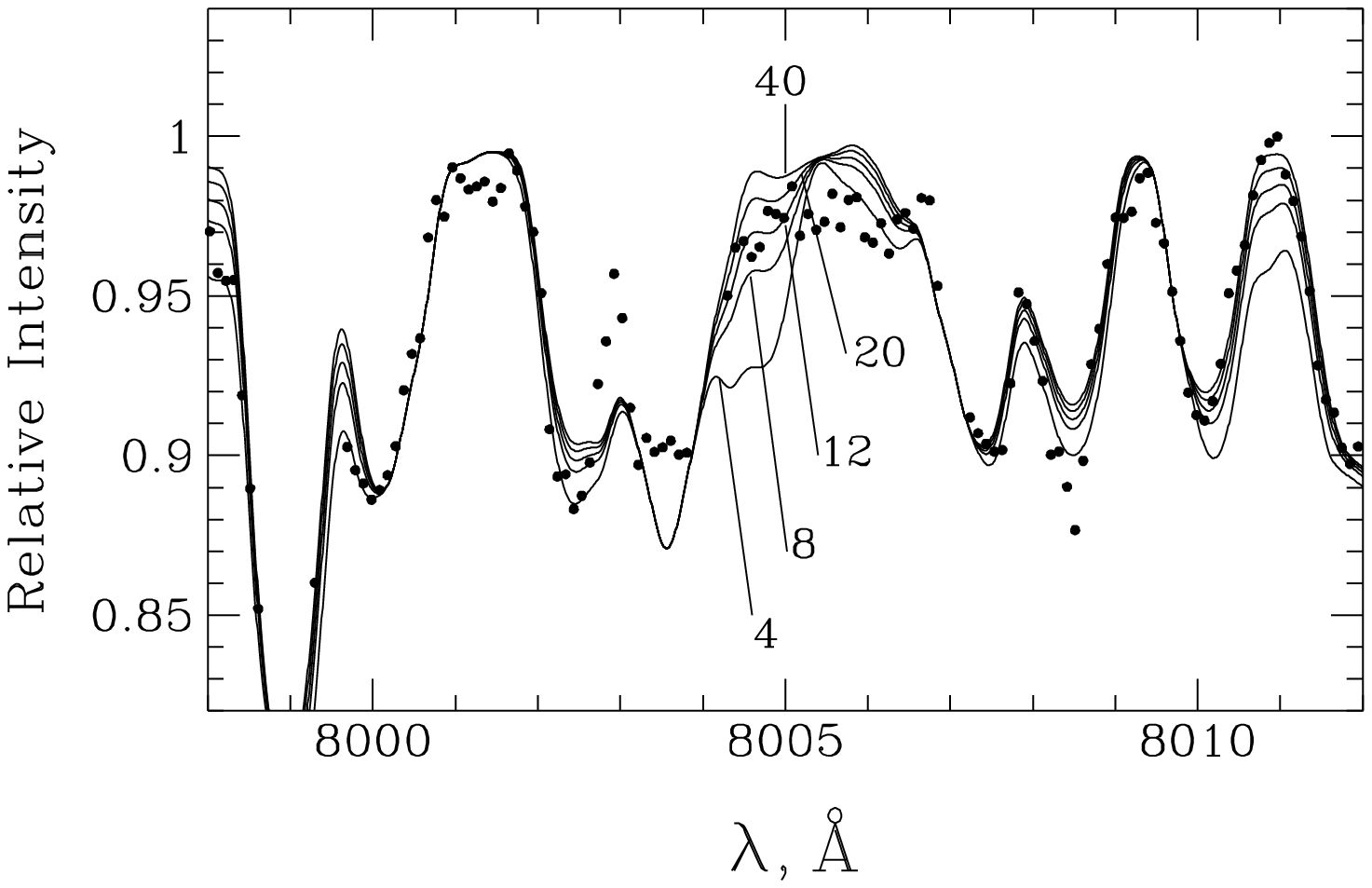]
{Star PDS\,365. Observed (points) and synthetic (lines) spectra
calculated with different values $^{12}$C/$^{13}{\rm C}= 4,$ 8, 12, 20, and 40,
${\rm [C/H]}=-0.1$, ${\rm [N/H]}=0.0$ and ${\rm [O/H]}=-0.04$. The best 
fit is achieved with   $^{12}$C/$^{13}{\rm C}\sim 12$.}

\figcaption[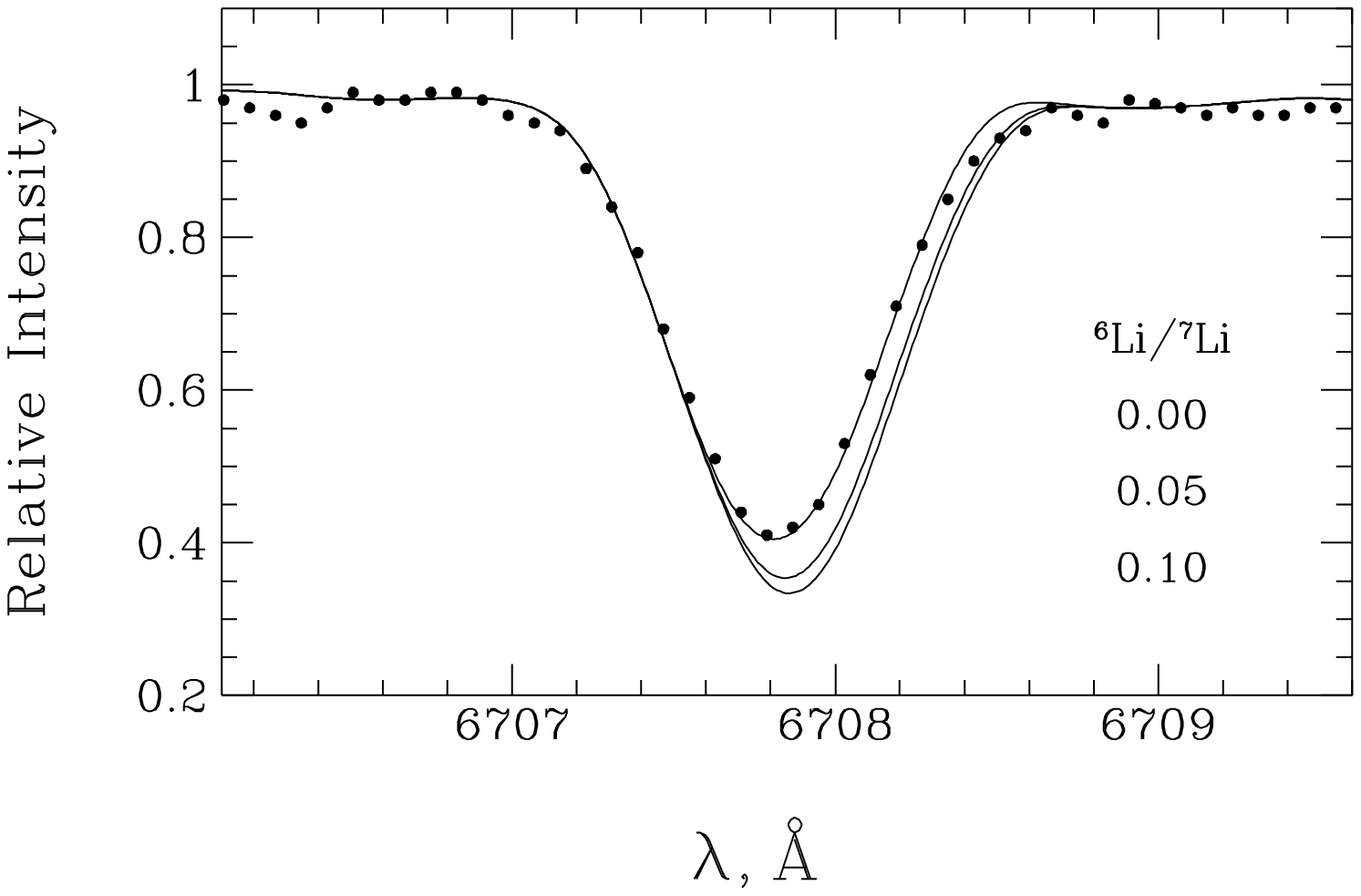]
 {Star PDS\,365. Calculations of the resonance Li\,{\sc i} line corresponding to the
total Li abundance $\log \varepsilon({\rm Li}) =3.3$
for different
$^6$Li/$^7$Li ratios equal to 0.0, 0.05 and 0.10.
The observed spectrum is best fitted with $^6$Li/$^7$Li = 0.0.}

\figcaption[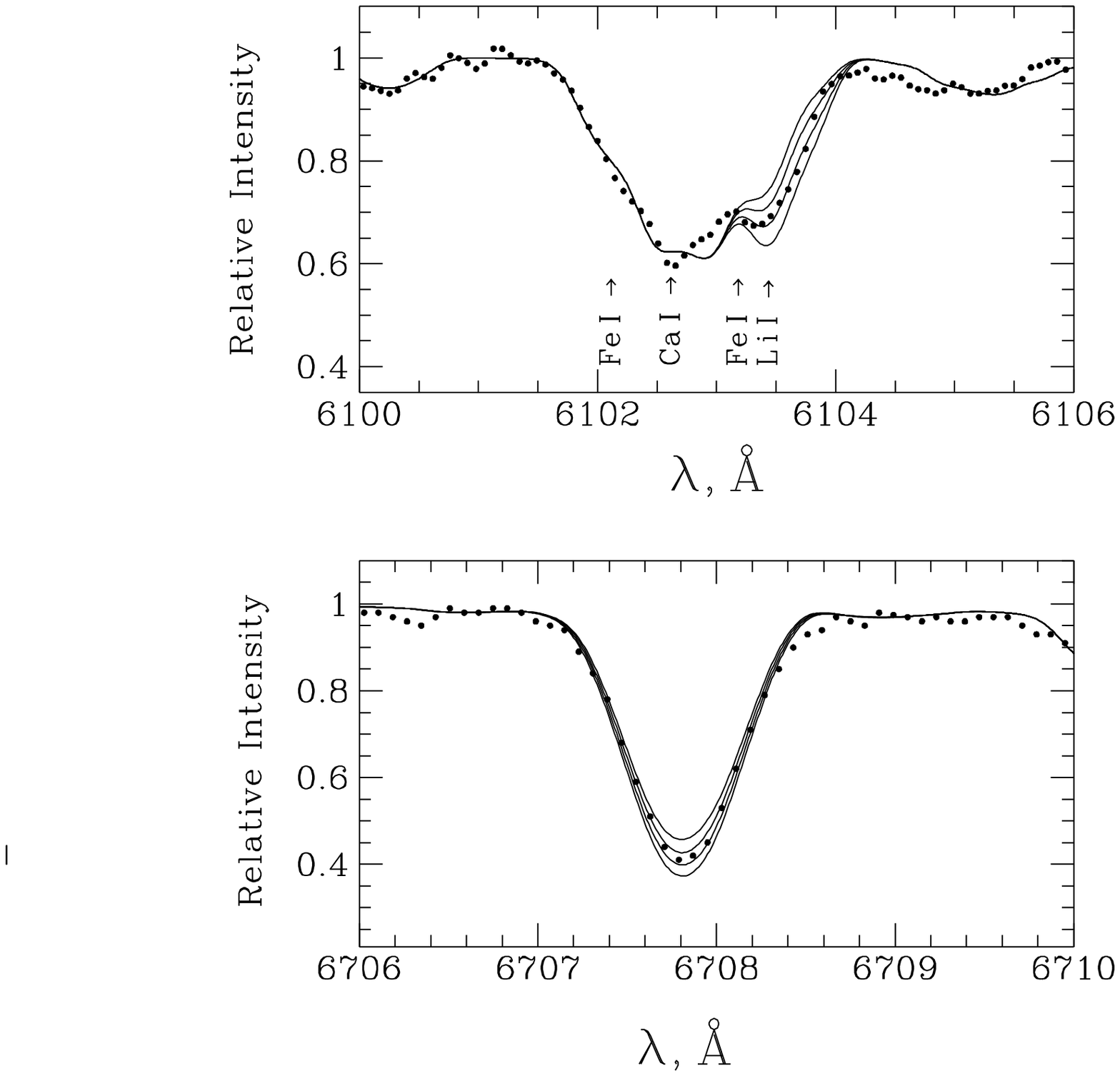]
{Star PDS\,365. Spectra  show the  Li\,{\sc i} features at
6708 \AA\ and 6104 \AA\ lines.
Observed profiles are represented by separated points.
Synthetic spectra are shown for four $^7$Li abundances equal to 2.9, 3.1, 3.3
and 3.5.
The best-fitting spectrum in both cases corresponds to  the  Li abundance
$\log\varepsilon{\rm (Li)} = 3.3$. }

\figcaption[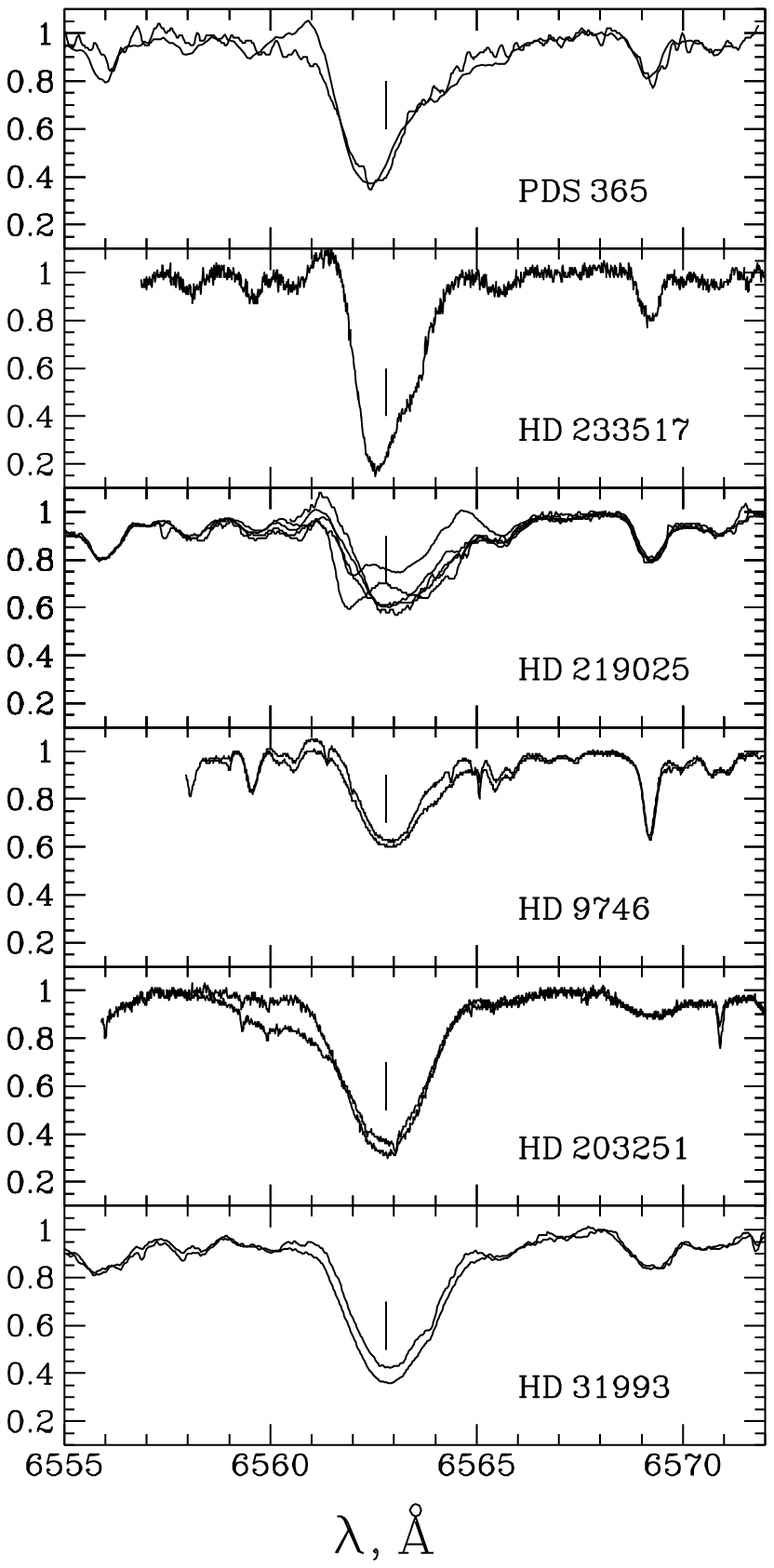]
{H${\alpha}$ profiles of some representative giants.
All spectra are presented on the same intensity scale, and corrected for stellar
radial velocity.
Vertical lines represent the heliocentric standard of rest. The H${\alpha}$ profile
for some stars have been obtained at different times spanning some years for PDS\,365 and
 HD\,219025 to
a few days for HD\,203251, HD\,9746, HD\,233517 and HD\,219025.}

\figcaption[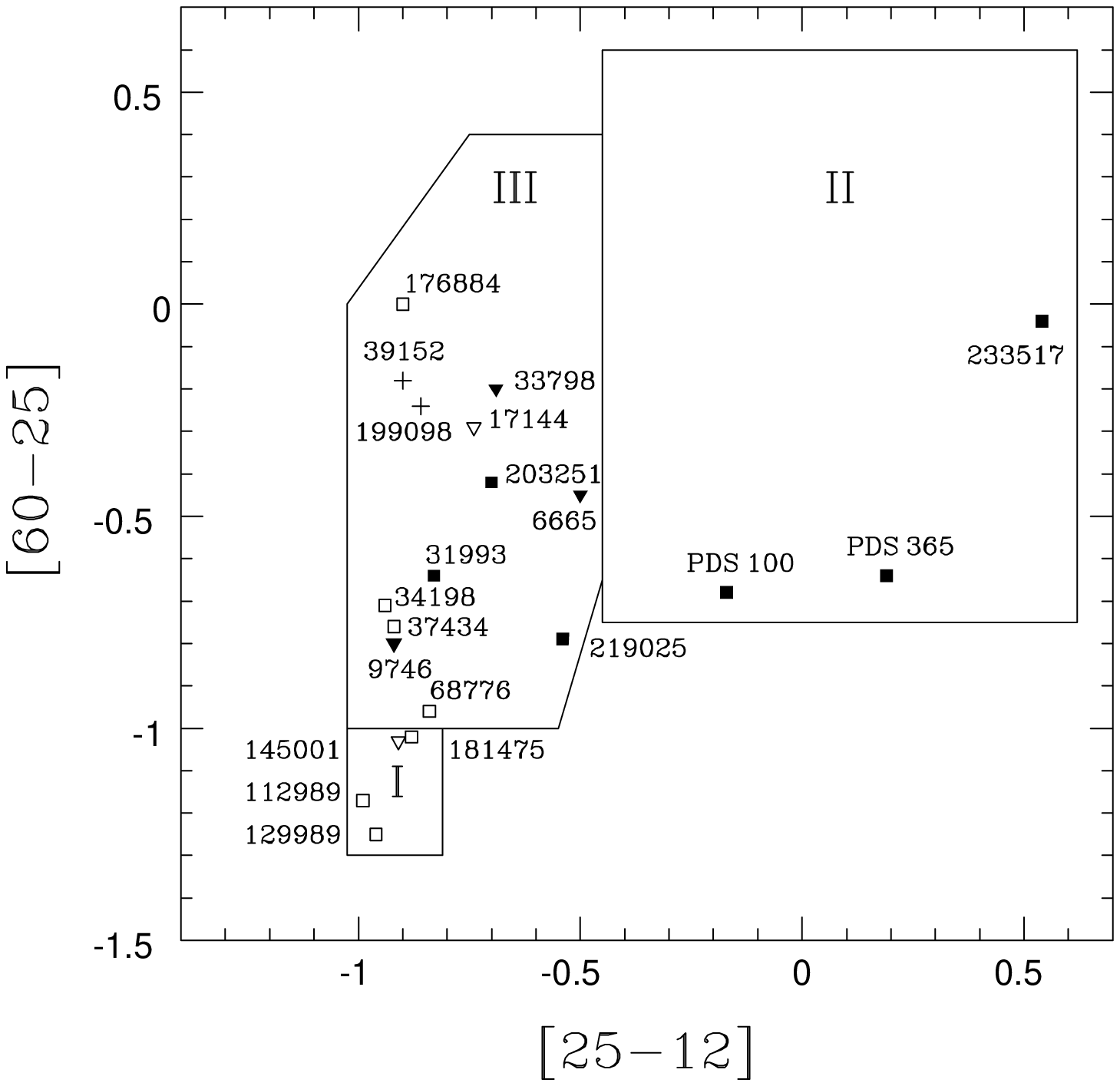 and 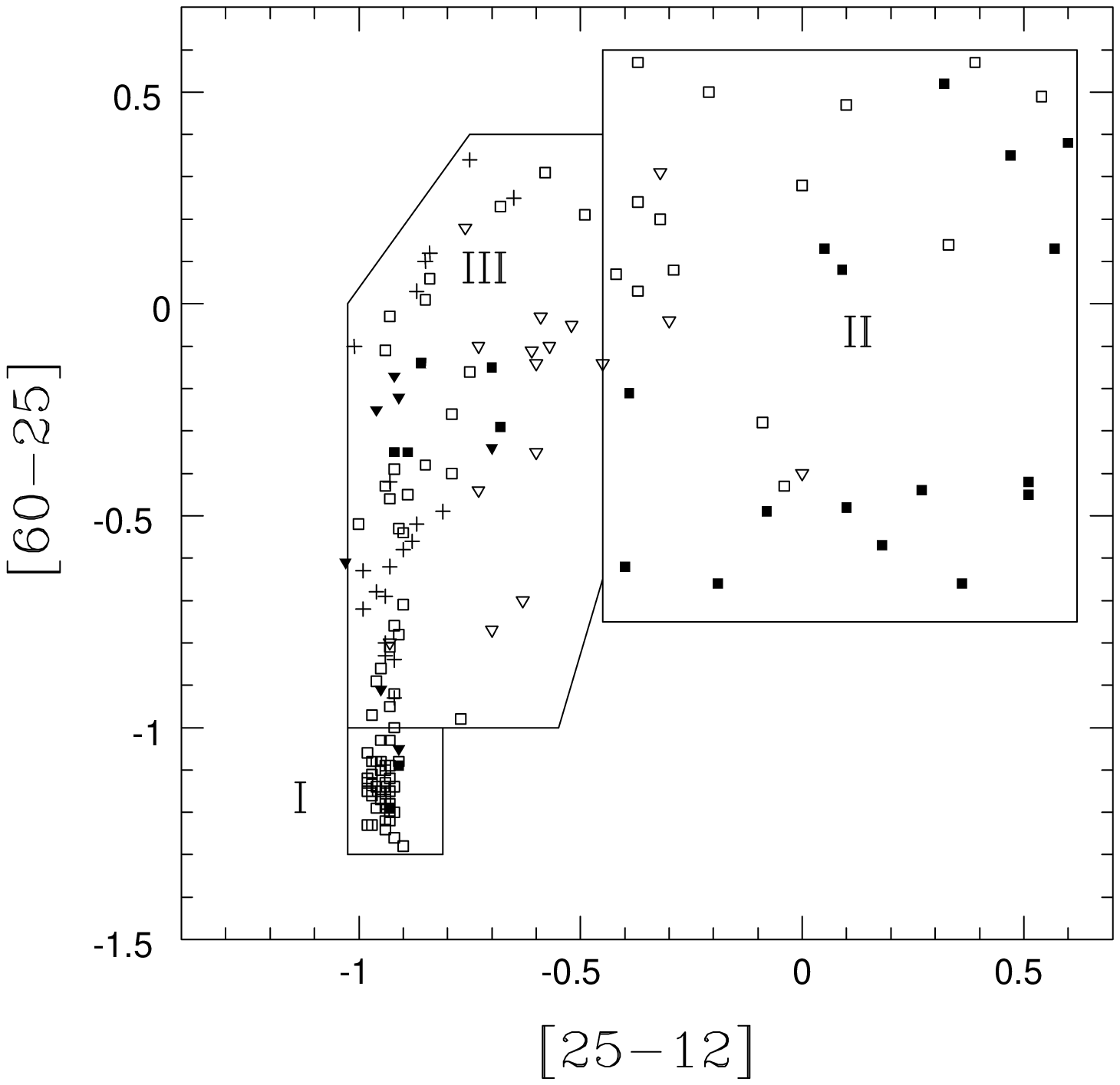]
{Distribution of the IRAS sources in a color-color diagram with
flux densities at 12, 25 and 60 microns for
 rapidly-rotating giants ({\it a}) and
for slowly-rotating giants ({\it b}).
Filled symbols
represent Li-rich objects. Triangles are
IRAS sources with one or two flux limits. Open symbols correspond to
Li-poor giants. Crosses -- giants for which no observations in the Li 
line have been made.}


\begin{thebibliography}{}

\bibitem[Alexander(1967)]{ale67} Alexander, J.B. 1967, The Observatory, 87, 238

\bibitem[Ambruster et al. (1997)]{amb97} Ambruster, C. W., Fekel, F.C., Guinan, E.F.,
            \& Hrivnak, B. J. 1997, \apj, 479, 960

\bibitem[Andersen et al.(1984)]{and84} Andersen, J.,  Gustafsson, B., \&  Lambert, D.L.
        1984,  \aap, 136, 65

\bibitem[Balachandran et al.(2000)]{bal2000} Balachandran, S.C., Fekel, F.C., Henry, G.W.,
        \&  Uitenbroek, H. 2000, \apj, 542, 978

\bibitem[Berdyugina \& Savanov(1994)]{ber94} Berdyugina, S.V., \& Savanov, I.S. 1994,
          Astron. Letters, 20, 639

\bibitem[Brown et al.(1989)]{bro89} Brown, J.A., Sneden, C., Lambert, D.L., \&
              Dutchover, E. 1989,  \apjs, 71, 293

\bibitem[Cameron \& Fowler (1971)] {cam71} Cameron, A.G.W., \& Fowler, W.A. 1971, 
        \apj, 164, 111

\bibitem[Carlsson et al.(1994)]{car99} Carlsson, M., Rutten, R.J., Bruls, J.H.M.J.,
        \& Shchukina, N.G. 1994, \aap,  288, 860

\bibitem[Castilho et al.(1999)]{cas99} Castilho, B. V., Spite, F., Barbuy, B.,
         Spite, M., de Medeiros, J. R., \& Gregorio-Hetem, J. 1999, \aap, 345, 249

\bibitem[Castilho et al.(2000)]{cas2000} Castilho, B. V., Gregorio-Hetem, J.,
        Spite, F., Barbuy, B., \& Spite, M. 2000, \aap, 364, 674

\bibitem[Charbonnel \& Balachandran(2000)]{cha2000} Charbonnel, C., \& 
         Balachandran, S.C. 2000, \aap, 359, 563

\bibitem[Claria(1980)]{cla80} Clari\'a, J.J. 1980, \apss, 72, 347

\bibitem[Cutispoto(1980)]{cut80} Cutispoto, G. 1995, \aaps, 111, 507

\bibitem[Davis \& Phillips(1963)]{dav63} Davis, S.P., \& Phillips, J.G. 1963,
           The Red System ($A^2\Pi - X^2\Sigma$) of the CN Molecule,
           University of California Press, Berkley and Los Angeles

\bibitem[de la Reza, \& da Silva(1995)]{rez95} de la Reza, R., \&
            da Silva, L. 1995, \apj, 439, 917

\bibitem[de la Reza, Drake, \& da Silva (1996)]{rez96} de la Reza, R., Drake, N.A., 
        \& da Silva, L. 1996, \apj, 456, L115

\bibitem[de la Reza et al.(1997)]{rez97} de la Reza, R., Drake, N.A., da Silva, L.,
        Torres, C.A.O., \& Martin, E.L. 1997, \apj, 482, L77

\bibitem[De Medeiros, Melo, \& Mayor(1996)]{med96} De Medeiros, J.R., Melo, C.H.F., \& 
         Mayor, M. 1996,  \aap, 309, 465

\bibitem[De Medeiros, Da Rocha, \& Mayor(1996)]{med96b} De Medeiros, J.R., Da Rocha, C.,
         \& Mayor, M. 1996, \aap, 314, 499

\bibitem[De Medeiros \& Mayor(1999)]{med99a} De Medeiros, J.R., \& Mayor, M. 1999, \aaps,
        139, 433

\bibitem[De Medeiros et al.(1999)]{med99b} De Medeiros, J.R., Konstantinova-Antova, R.K.,
                   \& Da Silva, J.R.P. 1999, \aap, 347, 550

\bibitem[De Medeiros et al.(2000)]{med20} De Medeiros, J. R., do Nascimento, J. D., Jr.,
         Sankarankutty, S., Costa, J. M., \& Maia, M.R.G. 2000, \aap, 363, 239

\bibitem[Den2000]{den2000} Denissenkov, P.A., \& Weiss, A. 2000, \aap, 358, L49

\bibitem[Drake(1998)]{dra98} Drake, N.A. 1998, PhD Thesis - ON

\bibitem[Eaton(1995)]{eat95} Eaton, J.A. 1995, \aj, 109, 1797

\bibitem[Efremov \& Sitnik(1988)]{efr88} Efremov, Yu.N., \& Sitnik, T.G. 1988, Sov.
                            Astron. Lett., 14, 347

\bibitem[Fekel(1988)]{fek88} Fekel, F.C. 1988,  A Decade of UV Astronomy with
         the IUE Satellite, vol.\,1, 331

\bibitem[Fekel \& Marschall (1991)] {fek91} Fekel, F.C. \& Marschall, L.A. 1991, \aj, 
         102, 1439

\bibitem[Fekel \& Balachandran(1993)]{fek93} Fekel, F.C., \& Balachandran, S. 1993,
      \apj, 403, 708

\bibitem[Fekel et al.(1996)]{fek96} Fekel, F.C., Webb, R.A., White, R.J.,
           \& Zuckerman, B. 1996, \apj, 462, L95

\bibitem[Fekel(1997)]{fek97} Fekel, F.C. 1997, \pasp, 109, 514

\bibitem[Fekel \& Watson(1998)]{fekwat19} Fekel, F.C., \& Watson, L.C. 1998, \aj, 116, 2466

\bibitem[Gondoin(1999)]{Gond99} Gondoin, P. 1999, \aap, 352, 217

\bibitem[Greg\'orio-Hetem et al.(1992)]{gre92} Greg\'orio-Hetem, J., L\'epine, J.R.D., Quast, G.R.,
              Torres, C.A.O., \& de la Reza, R. 1992, \aj, 103, 549

\bibitem[Greg\'orio-Hetem et al.(1993)]{gre93} Greg\'orio-Hetem, J., Castilho, B.V., 
              \& Barbuy, B. 1993, \aap, 268, L25

\bibitem[Henry et al. 1995]{hen95} Henry, G.W.,  Fekel, F.C., Hall, D.S.
         1995, \aj, 110, 2926

\bibitem[Henry et al. 2000]{hen2000} Henry, G.W., Fekel, F.C., Henry, S.M.,
         \& Hall, D.S.  2000, \apjs, 130, 201   

\bibitem[Hooten \& Hall (1990)]{hoo90} Hooten, J. T., \& Hall, D. S. 1990, \apjs, 74, 225

\bibitem[Houk \& Cowley (1975)]{houk75} Houk, N., \& Cowley, A.P. 1975, ``Michigan 
               Catalogue of two-dimensional spectral types  for the HD stars'', Vol. 4.,
              Ann Arbor: University of Michigan, Department of Astronomy

\bibitem[Jasniewicz et al.(1999)]{25} Jasniewicz, G., Parthasarathy, M., de Laverny, P.,
        \& Th\'evenin, F. 1999, \aap, 342, 831

\bibitem[Kj\ae rgaard et al. ]{kja82} Kj\ae rgaard, P., Gustafsson, B., Walker, G.A.H.,
       \& Hultqvist, L. 1982, \aap, 115, 145

\bibitem[Knauth et al. 2000]{kna20} Knauth, D. C., Federman, S. R., Lambert, D. L.,
          \& Crane, P. 2000, \nat, 405, 656

\bibitem[Konstantinova-Antova, \& Antov(2000)]{kon2000}
        Konstantinova-Antova, R., \& Antov, A. 2000, Kinematics and Physics of
             Celestial Bodies. Suppl. Ser. N3 , ``Astronomy in Ukraine - 2000
 and Beyond (Impact of International Cooperation)'', (ed. Ya.S. Yatskiv), p. 342

\bibitem[Konstantinova-Antova et al.(2000)]{kon-ant2000}
        Konstantinova-Antova, R.K., Antov, A.P., Bachev, R.S. 2000, 
        IVBS, 4867, 1

\bibitem[Kovacs (1969)]{kov69} Kovacs, I. 1969, {\it Rotational Structure in the Spectra
        of Diatomic Molecules}, Akademiai Kiado, Budapest

\bibitem[Kurucz (1992)]{kur92} Kurucz, R.L. 1992, Rev. Mex. Astron. Astrofis., 23, 45

\bibitem[Kurucz (1993)]{kur93} Kurucz, R. 1993, ATLAS9 Stellar Atmosphere
          Programs and 2 ${\rm km\,s^{-1}}$ Grid,
               {\rm Smithsonian Astroph. Obs.} CD-ROM no.$\;$13


\bibitem[Lambert (1978)] {lam78} Lambert, D.L. 1978, \mnras,  182, 249

\bibitem[Lambert \& Ries(1981)]{lambert81} Lambert, D.L., \& Ries, L.M. 1981, \apj, 248, 228

\bibitem[Lindg\"ard \& Nilsen(1977)]{lind77}      Lindg\"ard, A., \& Nielsen, S.E. 1977,
                 {\rm At. Data Nucl. Data Tables}, 19, 533


\bibitem[Lundstr\"om et al.(1991)]{lun91} Lundstr\"om, I., Ardeberg, A., 
          Maurice, E., \& Lindgren, H. 1991, \aaps,       91, 199

\bibitem[Mallik(1993)]{mal93}  Mallik, S.V. 1993, \apj, 402, 303

\bibitem[McWilliam \& Rich(1994)]{mcwil94} McWilliam, A., \& Rich, R.M. 1994, \apjs,
               91, 749

\bibitem[Mishenina \& Tsymbal(1997)]{mishe97} Mishenina, T.V., \& Tsymbal, V.V. 1997, 
         Astron. Letters, 23, 609

\bibitem[Moore 1966]{moo66} Moore, C.E., Minnaert, M.G.J., \& Houtgast, J. 1966,
              {\rm The Solar Spectrum from 2935 \AA$\;$ to 8770 \AA,} NBSM

\bibitem[Pagel]{pag95} Pagel, B.E.J., ${\rm Tautvai\check siene}$, G., 1995,
                  \mnras, 276, 505

\bibitem[Palacios et al.(2001)]{pal01} Palacios, A., Charbonnel, C, Forestini, M. 2001, \aap, 375, 9

\bibitem[Pavlenko et al.(1999)]{pav99} Pavlenko, Ya.V., Savanov, I.S., Yakovina, L.A.
                   1999,  Astron. Rep. 43, 671

\bibitem[Percy et al.(1994)]{per94} Percy, J.R., Wong, N., Bohme, D., et al.
            1994, \pasp, 106, 611

\bibitem[Phillips\& Davis (1968)] {phi68} Phillips, J. G., \& Davis, S.P. 1968,
         {\it The Swan System of the C$_2$ Molecule}, University of California
          Press, Berkeley and Los Angeles

\bibitem[Randich 1993]{ran93} Randich, S., Gratton, R., \& Pallavicini, R. 1993, \aap, 273, 194

\bibitem[Randich 1994]{ran94} Randich, S., Giampapa, M.S., \& Pallavicini, R.  1994,
                       \aap, 283, 893

\bibitem[Reddy et al.]{red02} Reddy, B. E., Lambert, D.L., Hrivnak, B. J., \& Bakker, E. J. 2002, \aj, in press


\bibitem[Sackmann \& Boothroyd(1999)]{sac99} Sackmann, I.-J., \& Boothroyd, A.I. 1999, \apj, 510, 217

\bibitem[Sackmann \& Boothroyd(2000)]{sac20} Sackmann, I.-J., \& Boothroyd, A.I.
             2000, in ``The Light Elements
                    and their Evolution'' IAU Symp. 198, Natal - Brazil, p. 98

\bibitem[Schmidt-Kaler(1982)]{sch82} Schmidt-Kaler, Th., 1982, in Landolt-B\"ornstein: NS,
               Group VI

\bibitem[Siess \& Livio(1999a)]{sie99a} Siess, L., \& Livio, M. 1999a, \mnras, 304, 925

\bibitem[Siess \& Livio(1999b)]{sie99b} Siess, L., \& Livio, M. 1999b, \mnras, 308, 1133

\bibitem[Skinner et al.(1995)]{skin95} Skinner, C.J., Sylvester, R.J., Graham, J.R., 
                  Barlow, M.J., Meixner, M., et al.  1995, \apj, 444, 861

\bibitem[Smith \& Shetrone(2000)]{2000} Smith, G.H., \& Shetrone, M.D. 2000, \pasp, 112, 1320

\bibitem[Sneden (1973)]{sne73} Sneden, C. 1973, \apj, 184, 839

\bibitem[Sneden \& Lambert (1982)] {sne82} Sneden, C., \& Lambert, D. L. 1982, \apj, 259, 381

\bibitem[Strassmeier et al.(1988)]{stra88} Strassmeier, K. G., \& Hall, D.S. 1988,
      \apjs,  67, 453

\bibitem[Strassmeier et al.(1997)]{stra97} Strassmeier, K. G., Bartus, J.,
         Cutispoto, G., \& Rodono, M. 1997, \aaps,  125, 11

\bibitem[Strassmeier et al.(2000)]{stra2000} Strassmeier, K., Washuettl, A.,
               Granzer, Th., Scheck, M., \& Weber, M. 2000, \aaps, 142, 275

\bibitem[Tomkin et al. (1976)]{tomkin76} Tomkin, J., Luck, R.E., \&  Lambert, D.L. 1976, 
          \apj, 210, 694

\bibitem[Torres et al. (1995)]{tor95} Torres, C.A.O., Quast, G., de la Reza, R.,
              Greg\'orio-Hetem, J., \& L\'epine, J.R.D. 1995, \aj,
              109, 2146

\bibitem[Torres (1998)]{tor98} Torres, C.A.O. 1998, Internal Publication - ON

\bibitem[Torres (2000)]{tor20} Torres, C.A.O., Quast, G., de la Reza, R.,
          \& da Silva, L. 2000, in ``The Light Elements
                    and their Evolution'' IAU Symp. 198, Natal - Brazil, p.320

\bibitem[Torres et al.(2002)]{tor02} Torres, C.A.O., Quast, G., de la Reza, R.,
       da Silva, L., L\'epine, J.R.D., \& Greg\'orio-Hetem, J. 2002, to be submitted
       to the \aj

\bibitem[Whitelock(1995)]{whi1995} Whitelock, P., Menzies, J., Feast, M.,  Catchpole, R., 
         Marang, F., \& Carter, B. 1995, \mnras, 276, 219

\bibitem[Wyller(1966)]{wyl66} Wyller, A.A. 1966, \apj, 143, 828

\end{thebibliography}
\end{document}


{\small 
\begin{deluxetable}{lcrlllc}
\tablecolumns{7}
\tablewidth {0pc}
\tablecaption {Fast rotating giants, spectral types,  magnitudes, rotation velocities, Li abundances and manifestations of activity} 
\tablehead{
\colhead{Star} & \colhead{Sp.~Type} & \colhead{$V\;\;$} & \colhead{$v\sin i$ (km\,s$^{-1}$)} & \colhead{EW(Li) (m\AA)}  & \colhead{$\log 
                                                                                      \varepsilon({\rm Li}) $} & \colhead{Activity} \\
}
\startdata
BD+25\,4819 &    K0\,II    &  10.94     &  $14.9^{10}$         & $<\!10^{19}$ &             &     \\  
BD+31$\,$2471 &    K0\,II  & $10.9\;\,$ &  $13.8^{10}$         &            &             &     \\   
BD+70$\,$959  &    K0\,III  & 9.64      &  $23^{7}$         &  $90^{7}$  &   $1.8^2$  & X-ray, Ca\,II    \\ 
HD$\,$6665    &  G5\,III   & 8.44       & $10^{14}$, $7.9^{14}$, $6.0^{19}$&$398^{14},\,405^{19}$            & $2.92^{14},\,2.7^{19}$     & Ca\,II   \\
HD$\,$9746    &    K1\,III &  6.22      & $8^2,\;8.7^{10}$        & $396^2,\;455^{13},\,518^{19}$ &$2.8^2,\,3.6^{12},\,3.75^{13},\,3.5^{19}$  & X-ray, flares, Ca\,II  \\ 
HD\,17144 &K1\,III/G8\,III-IV& 8.27     & $15^2,\;18.9^{10}$           & $55^2,\,85^{19}	$   &       $1.1^2,\,0.85^3$    & X-ray, Ca\,II    \\  

HD\,31993   &    K2\,III   & 7.49       & $33.4^{5},\;31.1^{10}$        & $144^2,\,174^{19}$  &       $1.4^2,\,1.0^3,\,1.7^{16}$    &  X-ray, Ca\,II   \\ 
HD\,33798   &    G8\,III   & 6.92       &  $33.1^{5}$              & $46^2$  & $1.5^2,\;1.8^1$    & X-ray, flares, Ca\,II \\ 
HD\,34198   &    K0\,III   &  6.95      &  $15^2,\;18.7^{10}$          & $13^2$  &  $0.4^2,\;<\!-0.2^3$    &  Ca\,II \\ 
HD\,37434   &    K2\,III   & 6.10       &  $46^2,\;65^{10}$            &            &$<\!-0.15^3$    & X-ray, Ca\,II \\ 
HD\,39152   &    K0\,II    & 7.41       & $8.0^{10}$                   &            &                &     \\
HD\,68776   &    G8\,III   &  6.37      & $8.9^{10}$                   &           &  $1.1^{19}$            &          \\ 
HD\,112989  &  G9\,III\,CH-2F& 4.90     & $11.0^{10},\;12^{15}$        &             & $+0.9^1$     & X-ray, flares, Ca\,II \\ 
HD\,129989  & K0\,II-III   & 2.70       &  $8.4^{10}$                  &  $<\!10^{19}$ &              &  \\ 
HD\,145001  &    G8\,III  &5.00         &  $9.9^{10}$                  &             & $<\!+0.5^1$  & X-ray  \\ 
HD\,176884  &    K0\,III  &  6.01       & $15.0^{11}$               &             &   $1.2^{11}$    &   \\ 
HD\,181475  &K5\,II, K7\,IIa&6.96       & $8.2^5,\,8.4^{10}$                   &   $<\!10^{19}$ &           & Ca\,II  \\ 
HD\,185958  & G8\,IIIaCN\,0.5& 4.39     & $9.9^{10}$                &             & $<\!0.3^1,\,0.2^6$  &   \\ 
HD\,199098  &    K0\,II   &   5.47      & $10.7^{10}$                  &             &              &        \\ 
 HD\,203251   &    K2\,III  &  8.03     &  $40^2,\;44.8^{10}$          & $214^{19}$     & $1.4^2,\,1.5^{19}$   &  Ca\,II  \\  
 HD\,219025  &    K2\,III  &  7.68      & $20^3,\,23^{11}$          &  $456^{9}$, $428^{19}$ & $3.3^{9},\,3.0^{11},\,2.9^{19}$  &  Ca\,II    \\     
 HD\,232862  &    G8\,II   &   9.46     & $20.6^{10}$                  &             &               &   X-ray      \\ 
 HD\,233517  &  K2\,III/II & $9.4\;\,$  &  $17.6^{13}$              &$550^8,\;546^{13}$&  $3.9^8,\,4.22^{13}$  & Ca\,II \\ 
 HD\,284857  &   K2\,II    & 9.30       & $8.0^{10},\,4.5^{19}$        &   $43^{19}$    & $0.3^{19}$            &    \\  
%
PDS\,100    &    K0-2\,III  & 10.45     & $9^{17}$                  & $372^{17}$  &    $2.5^{17}$  &   Ca\,II  \\
PDS\,365    &    K1\,III  & 13.15       & $20^{19}$                 & $450^{19}$, 465$^{18}$  &    $3.3^{19}$  &  Ca\,II   \\  
\enddata
\tablerefs{$^1$Brown et al. 1989, 
               $^2$Fekel \& Balachandran 1993, 
               $^3$Randich et al. 1993,
               $^4$Fekel et al. 1996, 
               $^5$Fekel 1997,
               $^6$Mishenina \& Tsymbal 1997,
               $^7$Ambruster et al. 1997,
               $^8$Drake 1998,
               $^9$Fekel \& Watson 1998, 
               $^{10}$de Medeiros \& Mayor 1999, 
               $^{11}$Jasniewicz et al. 1999, 
               $^{12}$Pavlenko et al. 1999, 
               $^{13}$Balachandran et al. 2000, 
               $^{14}$Strassmeier et al. 2000,
               $^{15}$Henry et al. 2000,
               $^{16}$Castilho et al. 2000,
               $^{17}$Reddy et al. 2002,
               $^{18}$Torres et al. 2002,
               $^{19}$This work. }
\tablecomments{Lithium abundances in references 12 and 13  are in non-LTE.}
\end{deluxetable}

\clearpage

\begin{deluxetable}{lcccc}
\tablecolumns{5}
\tablewidth {0pc}
\tablecaption {}
\tablehead{
\colhead{Star} & \colhead{Telescope}  & \colhead{Spectrograph}   & \colhead{Spectral Region}  & \colhead{Date}   \\
}
\startdata
PDS\,365     &  CTIO, 4.0m            & Coude                 & 5120 \AA\ -- 8370 \AA  & 05/01/96    \\
PDS\,365     &  ESO, 1.52m            & FEROS                 & 3800 \AA\ -- 9000 \AA  & 06/05/01    \\ 
BD+25\,4819  &  ESO, 1.52m            & FEROS                 & 3800 \AA\ -- 9000 \AA  & 09/09/01    \\
HD\,6665     &  ESO, 1.52m            & FEROS                 & 3800 \AA\ -- 9000 \AA & 12/10/00   \\  
HD\,9746     & McDonald, 2.7m         & Coude                 & 5680 \AA\ -- 8150 \AA, with gaps & 11/05/97    \\
             &                        &                       &                                  & 11/09/97    \\  
HD\,17144    &  ESO, 1.52m            & FEROS                 & 3800 \AA\ -- 9000 \AA         & 12/05/00    \\

HD\,31993    & ESO, 1.52m             & FEROS                 & 3800 \AA\ -- 9000 \AA         & 12/05/00    \\
             &                        &                       &                               & 12/07/00    \\
             &                        &                       &                               & 12/09/00    \\
HD\,68776    & ESO, 1.52m             & FEROS                 & 3800 \AA\ -- 9000 \AA         & 12/06/00    \\
HD\,129989   & ESO, 1.52m             & FEROS                 & 3800 \AA\ -- 9000 \AA         & 06/02/01    \\
HD\,181475   & ESO, 1.52m             & FEROS                 & 3800 \AA\ -- 9000 \AA         & 06/02/01    \\
HD\,203251   & McDonald, 2.7m         &  Coude                & 5680 \AA\ -- 8150 \AA, with gaps  & 11/05/97  \\
             &                        &                       &                               & 11/09/97    \\
HD\,233517   & McDonald, 2.7m         & Coude                 & 5680 \AA\ -- 8150 \AA, with gaps & 11/06/97    \\ 
             &                        &                       &                               & 11/08/97    \\
HD\,219025   &  ESO, 1.40m            & CAT CES Multichannel  & 5890 \AA, 6700 \AA            &   09/18/98   \\
             &  ESO, 1.52m            & FEROS                 & 3800 \AA\ -- 9000 \AA & 22/06/99    \\
             &                        &                       &                               & 12/23/99    \\
             &                        &                       &                               & 07/23/00    \\
             &                        &                       &                               & 08/21/00    \\
             &                        &                       &                               & 10/08/00    \\
             &                        &                       &                               & 12/14/00    \\
             &                        &                       &                               & 12/15/00    \\
             &                        &                       &                               & 12/16/00    \\
             &                        &                       &                               & 01/31/01    \\
HD\,284857   &  ESO, 1.52m            & FEROS                 & 3800 \AA\ -- 9000 \AA         & 12/05/00    \\ 
$\epsilon$ Vir &  ESO, 1.52m          & FEROS                 & 3800 \AA\ -- 9000 \AA         & 08/12/01   \\ 
\enddata
\end{deluxetable}

\clearpage

\begin{deluxetable}{lcccrr}
\tablecolumns{6}
\tablewidth {0pc}
\tablecaption {}
\tablehead{
\colhead{Ion} & \colhead{$\lambda,\,$\AA}& \colhead{$\chi,\,$}& \colhead{$\log gf$} & \colhead{EW$_{\rm obs}$} & \colhead{[Fe/H]}\\
\colhead{}    & \colhead{}               & \colhead{eV}       & \colhead{}          & \colhead{{\rm m\AA}} &   \\
}
\startdata
Fe\,I &   6151.623   &  2.18   &  $-3.33$   &  122.7   &  $-0.19$    \\  
      &   6173.341   &  2.22   &  $-2.86$   &  144.6   &  $-0.19$    \\ 
      &   6180.209   &  2.73   &  $-2.65$   &  135.1   &    0.07    \\ 
      &   6229.232   &  2.83   &  $-2.82$   &  105.3   &  $-0.20$    \\ 
      &   6271.283   &  3.32   &  $-2.80$   &   91.8   &    0.15    \\ 
      &   6297.799   &  2.22   &  $-2.66$   &  162.6   &  $-0.08$    \\ 
      &   6322.694   &  2.59   &  $-2.43$   &  141.1   &  $-0.23$    \\ 
      &   6335.337   &  2.20   &  $-2.17$   &  197.7   &  $-0.09$    \\ 
      &   6336.830   &  3.69   &  $-0.66$   &  167.9   &  $-0.19$    \\ 
      &   6353.849   &  0.91   &  $-6.36$   &   54.9   &    0.03    \\ 
      &   6355.035   &  2.84   &  $-2.32$   &  149.9   &    0.13    \\ 
      &   6392.538   &  2.28   &  $-3.95$   &   85.3   &  $-0.15$    \\ 
      &   6475.632   &  2.56   &  $-2.92$   &  128.4   &  $-0.05$    \\ 
      &   6481.878   &  2.28   &  $-2.98$   &  134.6   &  $-0.23$    \\ 
      &   6574.254   &  0.99   &  $-5.00$   &  126.1   &  $-0.08$    \\ 
      &   6593.874   &  2.43   &  $-2.37$   &  175.5   &  $\;0.05$    \\ 
      &   6627.560   &  4.53   &  $-1.55$   &   63.0   &  $-0.08$    \\ 
      &   6646.966   &  2.60   &  $-3.91$   &   60.8   &  $-0.18$    \\ 
      &   6653.911   &  4.14   &  $-2.44$   &   40.4   &  $-0.06$    \\ 
      &   6703.576   &  2.76   &  $-3.05$   &  100.1   &  $-0.20$    \\ 
      &   6705.105   &  4.61   &  $-1.06$   &   84.1   &  $-0.12$    \\ 
Fe\,II &  6416.928   &  3.89   &  $-2.68$   &   47.5   &  $-0.13$    \\
      &  6432.683    &  2.89   &  $-3.55$   &   66.9   &  $-0.06$    \\
      &  6456.391    &  3.90   &  $-2.28$   &   68.1   &  $-0.08$    \\
 
\enddata
\end{deluxetable}

\clearpage

\begin{deluxetable}{ll|ll|ll}
\tablecolumns{6}
\tablewidth {0pc}
\tablecaption {Main parameters of PDS\,365}
\tablehead{
\colhead{Parameter} & \colhead{Value}  & \colhead{Parameter} & \colhead{Value} & \colhead{Parameter} & \colhead{Value} \\
}
\startdata
$T_{\rm eff}$       & $4540\pm 150\;$K  & $(B-V)_{\rm obs}$   & $1.50^{\rm m}$  & [Fe/H]               & $-0.09\pm 0.05$     \\
$\log g$            & $2.20\pm 0.3$     & $d$                 & 1.6 kpc         &  [C/H]                 & $-0.1$      \\
$\xi_{\rm m}$       &$1.8\pm 0.2$ km\,s$^{-1}$ &$V_{\rm rad}$ & $-38.3\pm0.5$ km\,s$^{-1}$ & [N/H]       & $\;\;0.0$ \\
$L$                 &$72L_\odot$        & $v\sin i$ & $20.0\pm 1.0$ km\,s$^{-1}$  & [O/H]                  & $-0.04$        \\
$M$                 &$1.1 M_\odot$      &                     &                 & $^{12}$C/$^{13}$C    & $12^{-2}\!\!\!\!\!\!_{+8}$ \\
$R$                 & $14 R_\odot$      &                     &                 & $\log \varepsilon {\rm (Li)}$ & $3.3\pm 0.2$  \\
\enddata
\tablecomments{Due to the absence of Hipparcos parallax measurements of PDS\,365, $L$, $M$, $R$ and $d$ have only 
          approximate values. 
          This is also the case of [O/H] due to the presence of night sky emission at the [O\,{\sc i}] 6300 \AA\ line.}
\end{deluxetable}

\clearpage

\begin{deluxetable}{lcccccc}
\tablecolumns{5}
\tablewidth {0pc}
\tablecaption {Comparative CNO abundances}
\tablehead{
\colhead{Star} &\colhead{[C/Fe]} & \colhead{[N/Fe]} & \colhead{[O/Fe]} & \colhead{[Fe/H]} & \colhead{$^{12}$C/$^{13}$C} &
                                      \colhead{($T_{\rm eff}$, $\log g$)}  \\
}
\startdata
PDS\,365$^5$   & $-0.01$         & $+0.09$             & $+0.05$       & $-0.09$          & 12   & (4540, 2.2)  \\
PDS\,100$^1$   & $-0.07$         & $+0.06$             & $-0.05$       & $+0.14$          & \,9   & (4500, 2.5)  \\
HD\,9746$^2$   & $-0.08$         & $-0.16$             & $-0.05$       & $-0.13$          & 24   & (4420, 2.3)  \\
$\epsilon$ Vir$^5$ & $-0.25$     & $+0.19$             & $-0.35$       & $\,+0.05^a$      & 20    & (5000, 2.5)  \\
$\epsilon$ Vir$^3$ & $-0.35$     & $+0.09$             & $-0.28$       & $+0.05$          & $\,20^4$ & (5060, 2.7)  \\
\enddata
\tablerefs{$^1$Reddy et al. 2002, $^2$Berdyugina \& Savanov (1994), $^3$Kj\ae rgaard et al. (1982),
          $^4$ Lambert \& Ries (1981), $^5$This work.}
\tablecomments{
The solar abundances used in (2) and (5) are $\log \varepsilon{\rm (C_\odot)}=8.56$,
          $\log \varepsilon{\rm (N_\odot)}=8.05$, $\log \varepsilon{\rm (O_\odot)}=8.93$, and 
          $\log \varepsilon{\rm (Fe_\odot)}=7.50$. 
}
\tablecomments{
                  $^a$[Fe/H] adopted from (3) 
}
\end{deluxetable}